\documentclass[11pt]{myclass}

\usepackage{amssymb,amsmath,amsfonts,pdfpages,color, url}
\usepackage{booktabs}
\usepackage{wrapfig}
\usepackage{amsmath}
\usepackage{multirow}

\def\argmin{\mathop{\rm argmin}}

\def\T{\mathop{\mathcal T}}
\def\clf{\mathop{\mathtt{clf}}}

\def\sklearn{\mathop{\mathtt{sklearn}}}
\def\tslearn{\mathop{\mathtt{tslearn}}}

\newcommand{\dQ}{\ensuremath{\mathtt{d}_{Q}}}

\begin{document}

\title{Classifying Spatial Trajectories}

\author{Hasan Pourmahmood-Aghababa and Jeff M. Phillips \\ University of Utah \\ \texttt{h.pourmahmoodaghababa@utah.edu}, \texttt{jeffp@cs.utah.edu}}




   

\keywords{Computational geometry, Machine learning, Classification, Trajectory mining}

\maketitle

\begin{abstract}
We provide the first comprehensive study on how to classify trajectories using only their spatial representations, measured on 5 real-world data sets.  Our comparison considers 20 distinct classifiers arising either as a KNN classifier of a popular distance, or as a more general type of classifier using a vectorized representation of each trajectory.  We additionally develop new methods for how to vectorize trajectories via a data-driven method to select the associated landmarks, and these methods prove among the most effective in our study.  These vectorized approaches are simple and efficient to use, and also provide state-of-the-art accuracy on an established transportation mode classification task.  In all, this study sets the standard for how to classify trajectories, including introducing new simple techniques to achieve these results, and sets a rigorous standard for the inevitable future study on this topic.
\end{abstract}

\let\thefootnote\relax\footnotetext{Jeff Phillips thanks his support from NSF CCF-1350888, CNS-1514520, CNS-1564287, IIS-1816149, CCF-2115677, and from Visa Research.}

\section{Introduction}

A trajectory is the tracing of an object through physical space, and is now a first-class object in spatial data analysis due to the ease of creating these objects via GPS trackers.  Also, their analysis has been at the forefront of geo-spatial research in clustering~\cite{BGLR15,BDGHKL18,DKS16,buchin2019klcluster,ZKT2006},  similarity measures~\cite{ACKGS2018,Cut2011,MSMB2015,AKW04}, classification~\cite{FSFMJC2021,GCMM2006,LH2013,LLK2019,STK2002,SVSA2010,XZLW2011,ZGTZZ2018,GZZTXZ2017,SCM2017,SCRDKW2014, PSHL2012,LHLC2011,DWF2009,LHLG2008, MP2022}, and transportation mode detection~\cite{DH2018,DCHR2020,ETNK2016,FLFCHLT2016,WLJL2017,ZLCXM2008,ZX2008,EJM2018,Var2015,TVE2014}.
In this paper we do not factor in the absolute time these traces are made, but mostly treat the trajectories as geometric objects in the plane.  Moreover, in this setting, we focus on the central machine learning tasks of being able to classify trajectories to predict if they belong to one of two classes.  

To instantiate what is spatial trajectory classification and why we need to do classification, consider traffic data.  Discovering driving behavior patterns are critical for any government (as part of infrastructure design), auto companies (to adapt products to future use cases), and auto insurance companies (to adjust rates for safe versus reckless behavior). 
The simplest version of this is classifying drivers or routes as part of a favorable/standard versus unfavorable/divergent class.  We posit that most driver analysis will either directly rely on such a task, or build on it as part of larger model.  
For instance, consider an adversary that uploads some corrupted trajectories in the road network database of a government. Then the need for a discriminator is a must. 
In many cases the spatial classifiers are used with the addition of extra metadata information about the vehicle, timing, or the person generating that route. The choice of which ones to use is very situation-dependent and such we do not focus on this.  However, we observe that the methods we recommend are easily adjustable to adding these information to the classifiers, as we will demonstrate in two of our experiments.

This is a fundamental task in spatial data analysis, but as far as we are aware, has not comprehensively been studied as a stand-alone challenge.  
Previous trajectory classification work includes \cite{PT19a} which constructs some experiments similar to ours to demonstrate where their proposed technique is effective; we compare to and build on this work, but aim for a more objective and comprehensive comparative study.  Other work considers subtrajectory~\cite{FSFMJC2021,DMMZ2022} or segment~\cite{GCMM2006} classification but is not applicable to our tasks.   For instance, \cite{FSFMJC2021} builds a recurrent neural net (RNN) tuned to specific properties of their data sets -- it centrally uses an encoded location id of each waypoint not available in general trajectory data sets, like the ones we consider.  The focus of \cite{GCMM2006} is on mode-of-flight identification from very short subsets of trajectories.   
Another line of work is on inferring transportation modes~\cite{DH2018,DCHR2020,ETNK2016,FLFCHLT2016,WLJL2017,ZLCXM2008,ZX2008,EJM2018,Var2015,TVE2014} which we directly compare against in Section \ref{sec: Geolife}.  
Yet, the core trajectory classification task factors into many important challenges, and is destined to have an ever-expanding role as spatial data analysis increases in automation.  For instance, given a GPS trace, can we determine which individual most likely made it, or what the mode of transportation was. Or does the trajectory represent favorable behavior (a healthy animal, a customer who will make a large purchase) or an unfavorable one (a diseased animal, a shoplifter). 

While in some cases, these full applications are pre-mature or should enact ethical safeguards before deploying, the task of building a classifier for trajectories will surely play an important role in their ultimate use.  As part of this project we have identified 5 data sets with at least two naturally occurring classes of trajectories, of which it makes sense to classify.  
These data sets are publicly available, and our methodology is simple and reproducible, and we hope that these evaluation tasks become benchmarks for future development in this area.  

Moreover, by using known techniques to combine distances meant to model trajectories, and other trajectory representations, the literature provides for dozens of ways classifiers can be built.  In particular, a distance between trajectories (e.g., dynamic time warping, discrete Fr\'echet distance) can be used within a KNN classifier.  In addition, methods for featurizing a trajectory~\cite{PT19a,PP21} into a vector representation can be paired with off-the-shelf classifiers, from say sk-learn.  Thus we explore and evaluate a large cross-section of classifiers in this paper seeking to provide guidance on which ones work the best.

Furthermore, we adapt and extend several of these techniques to produce new methods which are among the best performing in the benchmarks we have designed.  These methods include a new data-driven method to select the landmark points which the vectorized representations are derived from, and a voting mechanism to boost the effect of several classifiers.  
Moreover, we show these vectorized approaches can be combined with other features (e.g., velocity) and this achieves state-of-the-art performance on the standard task for predicting mode of transportation.

In all, this work initiates a formal study of classification tasks for spatial trajectories, identifies state-of-the-art methods -- some are developed as part of this work, and provides as set of benchmarks so this field can continue to evolve in a reproducible way.

\section{Preliminaries} \label{sec: vectorization}

In this section first we describe existing landmark-based feature mappings for curves that not only allow us to define a distance on curves but also enables using almost all machine learning algorithms on curves. 
Then we provide a couple of simple enhancements to these methods.  The first one is a data-driven method to select the landmark points based on which one is often causing a mistake, and the second uses multiple such classifiers built from different randomly chosen landmarks and returns the majority vote of all of the classifiers.  
We will see that both methods provide small but tangible improvements in classification results thereafter.  

\subsection{Curve Vectorization}

Formally, by a {\it curve} we mean the image of a continuous non-constant mapping $\gamma: [0,1] \to \mathbb{R}^2$. The set of all curves is denoted by $\Gamma$. We recall the notions of landmark-based feature mapping and distance which were introduced in  \cite{PT19a}. For any landmark $q \in \mathbb{R}^2$ one can consider the distance function $v_q: \Gamma \to \mathbb{R}$ defined by 
\[
v_q(\gamma) = {\rm dist}(q, \gamma) = \min_{p \in \gamma} \|q-p\|.
\]
Now for a set of landmarks $Q=\{q_1, \ldots, q_n\}$ we can vectorize the $v_q$ function to get a feature mapping $v_Q: \Gamma \to \mathbb{R}^n$ by 
\[
v_Q(\gamma) = (v_1(\gamma), \ldots, v_n(\gamma)),
\]
where $v_i(\gamma) = v_{q_i}(\gamma)$ for $1 \leq i \leq n$. Indeed, this contribution of each landmark are stacked to get an appropriate vector representation of curves which enables us to take advantage of machine learning models for trajectory datasets, makes some clustering (like K-means) 
trivial and K-nearest neighbor classifier very efficient using fast Euclidean near-neighbor libraries. 
Now the landmark-based distance $\dQ: \Gamma \times \Gamma \to \mathbb{R}$ is defined by 
\[
\dQ(\gamma, \gamma') = \frac{1}{\sqrt{n}} \|v_Q(\gamma) - v_Q(\gamma')\|,
\]
where $\| \cdot \|$ denotes the Euclidean distance in $\mathbb{R}^n$.   
We next consider 2 alternative ways to define the $v_q$ mapping, but for a set of landmarks $Q$ they create a vector and distance in an analogous way.  

The second feature mapping is $v_q^{\exp}$, which is a vectorization of the function 
\[
v_q^{\exp}(\gamma) = \exp\big(-v_q(\gamma)^2/\eta^2\big),
\]
and localizes the feature mapping $v_q$ to a neighborhood induced by a learned scale term $\eta$. 

We remark that, however, the feature mappings $v_Q$ and $v_Q^{\exp}$ and the distance $\dQ$ do not capture the orientation of curves, which encode the direction of the trajectories.  An orientation preserving version which extends $v_Q^{\exp}$ is  introduced in \cite{PP21}.  First assume we can define $n_p(q)$ the normal vector of a curve $\gamma$ (using the curve orientation to keep its direction consistent), where $p$ is the closest point on $\gamma$ to $q$, and a few other technical conditions~\cite{PP21}.  Assume $\Gamma'$ shows the set of all simple curves which are differentiable almost everywhere. Then define $v_q^{\varsigma}: \Gamma' \to \mathbb{R}$ for some scale parameter $\varsigma > 0$ by
\[
v_q^{\varsigma}(\gamma) = \langle n_{p}(q), q-p\rangle \exp\big(- \|q-p\|^2/\varsigma^2\big)/\varsigma.
\]
When $p$ is an endpoint (i.e. $p=\gamma(0)$ or $p=\gamma(1)$), we need a slightly different definition which keeps the values local and continuous
\[
v_q^{\varsigma}(\gamma) = \frac{1}{\varsigma} \langle n_p, \frac{q-p}{\|q-p\|} \rangle \|q\|_{\infty,p} \, \exp\big(- \|q-p\|^2/\varsigma^2\big),
\]
where $\|q\|_{\infty,p}$ is the $l^{\infty}$-norm of $q$ in the coordinate system with axis parallel to $n_p$ and tangent line at $p$ and origin at $p$.

\begin{figure}[ht]
\includegraphics[width=0.32 \textwidth]{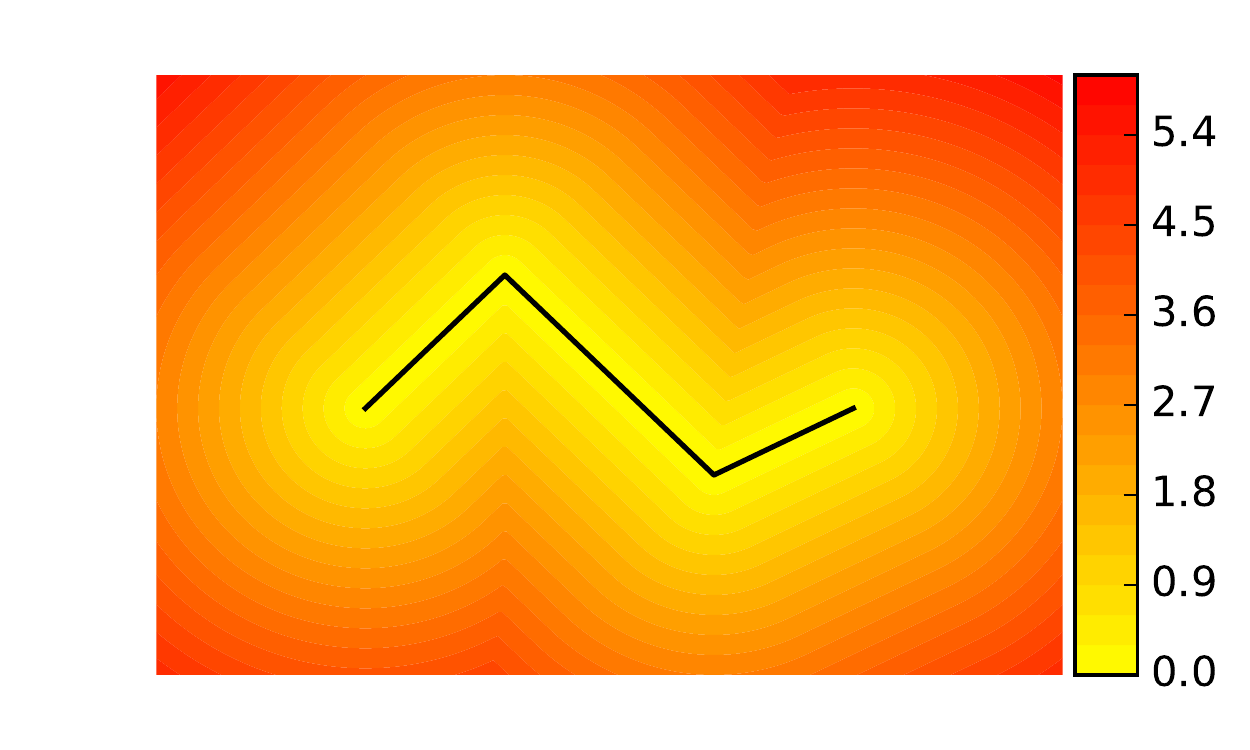}
\includegraphics[width=0.32 \textwidth]{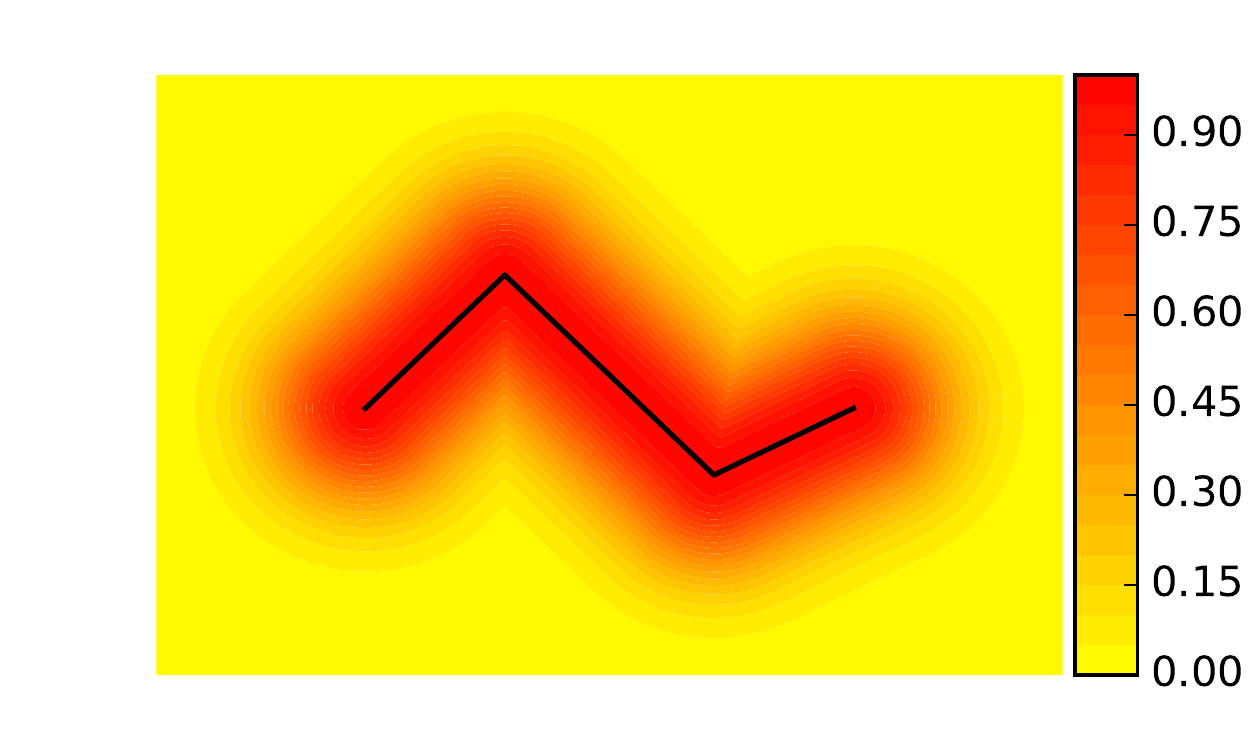}
\includegraphics[width=0.32 \textwidth]{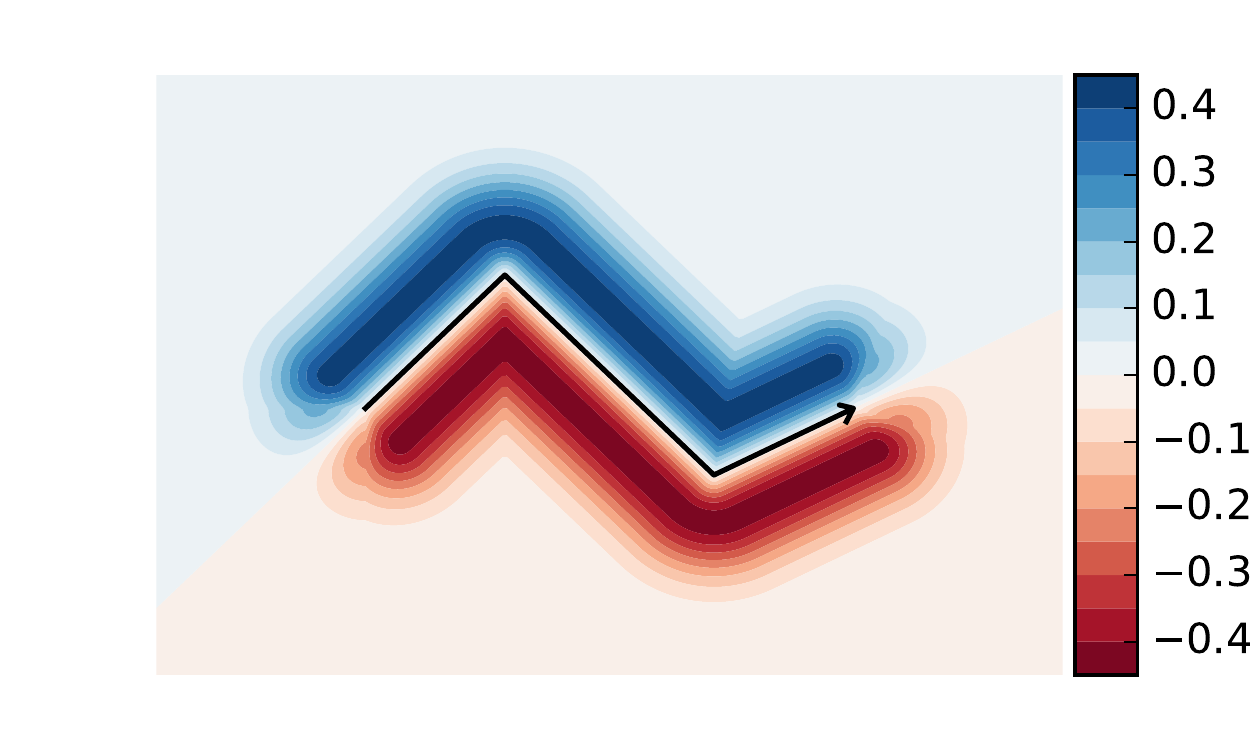}
\caption{Left: Example of feature mapping $v_q$ for a curve. Middle: Example of feature mapping $v_q^{\exp}$ for a curve ($\eta=2$). Right: Example of feature mapping $v_q^\varsigma$ for a curve, with sidedness encoded by positive/negative values ($\varsigma=1$).}  
\label{fig: curve2vec}
\end{figure}

Another landmark-based distance, denoted $\dQ^{\pi}$, was introduced in \cite{PT19a}, which we will evaluate via KNN classifier. Considering a landmark set $Q=\{q_1, \ldots, q_n\}$ and two curves $\gamma, \gamma'$ we can define 
$\dQ^{\pi}(\gamma, \gamma') = \frac{1}{n} \sum_{i=1}^n \| p_i - p_i' \|$,
where $p_i = \argmin_{p \in \gamma} \|q_i - p\|$ and $p_i' = \argmin_{p \in \gamma'} \|q_i - p\|$. Notice if there are multiple points available as argmin points, considering the natural parametrization of curves on the interval $[0,1]$ and thus a linear order on the curve, the first point will be chosen. For an illustration of feature mappings $v_q$, $v_q^{\exp}$, $v_q^{\varsigma}$ see Figure \ref{fig: curve2vec}.

\subsection{Mistake-driven Algorithm for Choosing Landmarks}

In previous work \cite{PT19a,PP21}, the landmarks were chosen randomly (from some bounding domain) for experimental evaluation, or sometimes on a fine grid within theoretical analysis.  In this section we propose a new and more data-driven mechanism to select the landmarks.  We refer to has a \emph{mistake-driven} approach since it selects points nearby trajectories for which a classifier makes a mistake on, reminiscent of the perceptron algorithm.

Consider two training trajectory data sets $\T_1$ (examples with label $1$) and $\T_2$ (examples with label $0$) and set $\T=\T_1 \cup \T_2$ (as the training set), and a particular classifier we denote $\clf$. Our goal here is to identify a set of landmarks that help to improve misclassification error rates for use with $\clf$. To accomplish this, first we designate a scale parameter, say $\eta$, which controls the standard deviation of a random process placing landmarks near curves.  We set $\eta = \|m_1 - m_2\|$, where $m_1$ and $m_2$ are the mean of $(x,y)$-coordinates of trajectories in the first and second training data sets respectively. Next, we choose a random landmark $q_0$ by adding a Gaussian noise with standard deviation of $\eta$ to a uniformly randomly chosen waypoint of a uniformly randomly chosen trajectory. Then we train a classifier of type $\clf$ on the mapped data $v_Q^{\exp}(\T, \eta)$, where $Q=\{q_0\}$. Then we choose the most misclassified trajectory, say $\gamma$, according to their scores given by the trained classifier and select a random landmark $q_1$ by adding a Gaussian noise with standard deviation of $\eta$ to a uniformly randomly chosen waypoint of $\gamma$ and append it to $Q$. We do the same process with $Q=\{q_0, q_1\}$ to choose another landmark. We repeat this process until we get the desired number of landmarks. The pseudo code is given in Algorithm \ref{algo: choose landmarks}. 
These landmarks are then used to generate a vector representation for trajectories, and a final $\clf$ classifier is trained.

\begin{algorithm}
\caption{Mistake-driven approach to choose landmarks}
\label{algo: choose landmarks} {\small
\begin{algorithmic}
\STATE \textbf{Input:} Two sets of trajectories $\T_1, \T_2$, $n$ (the number of needed landmarks), $\eta$ and a classifier $\clf$. 
\STATE \textbf{Output:} A set of landmarks $Q$ of size $n$.
\STATE Randomly select a trajectory $\gamma_0$ in $\T=\T_1 \cup \T_2$.
\STATE With standard deviation of $\eta$, select a random landmark $q_0$ near to a point of $\gamma_0$; initialize $Q = \{q_0\}$.  
\FOR {$i = 1, \ldots, n-1$}
\STATE Train a model using $\clf$ on the vectorized data $v_Q^{\exp}(\T, \eta)$.
\STATE Select a trajectory $\gamma_i$ that is most misclassified among $\T$.
\STATE With standard deviation of $\eta$, select a random landmark $q_i$ near to a point of $\gamma_i$ and append it to $Q$. 
\ENDFOR
\RETURN $Q$
\end{algorithmic}} 
\end{algorithm}

\section{Experimental Evaluation of Trajectories Classification Methods} \label{sec: experiments}

We have done experiments on 5 real world and 1 synthetic data sets which are explained in detail in the following subsections. Table \ref{table: data sets} shows an overview of these data sets, detailing the data size and also the size of a bounding box.  We also show the scale parameter $\varsigma$ used in experiments, we generally aim to set it about the same or a bit larger than the bounding box.  In the Characters, T-drive and Geolife data sets where the data spread changes over the same domain, we keep it fixed for all experiments in that data set type.

\begin{table}[h]
\caption{This table represents an overview of data sets used in the experiments in this paper. Here ``Size'' shows the number of selected trajectories after preprocessing step and ``L $\times$ W'' shows the length and width of the rectangular region containing all the sampled trajectories. The notation $\varsigma$ shows the scale parameter.} 
\label{table: data sets} 
\centering
\begin{tabular}{p{35mm}p{20mm}p{15mm}p{23mm}p{5mm}}
\toprule
	{\bf Dataset}			&	{\bf Pairs}		&   {\bf Size}	&	{\bf L $\times$ W}	&   {\bf $\varsigma$}   \\ \midrule 
	
	{\bf Car-Bus}			& 	--			  	    & 	76, 44	            &    $0.67 \times 0.61$         &   1 	\\ \midrule 
	
	{\bf Simulated Car-Bus}	&	--	    &	228, 220	        &  	 $0.67 \times 0.62$         &   1	 \\ \midrule 
	
	{\bf Two Persons}		&	--		    &	124, 89	            &	$30.53 \times 15.64$        &   100 \\ \midrule
	
	$\hspace{1mm}$ \newline \newline	 {\bf Characters}		&	
	 {\it u, w} \newline  {\it n, w} \newline  {\it n, u}  \newline  {\it b, c} \newline  {\it c, o}	&
		125, 131 \newline 	125, 130	\newline	131, 130	\newline	174, 141	\newline	141, 142	&
		$112.63 \times 80.55$ \newline		$121.38 \times 87.84$ \newline		$124.82 \times 84.34$ \newline		$83.11 \times 82.55$	 \newline	$98.75 \times 61.08$    &   
		100 \newline 100 \newline 100 \newline 100 \newline 100 \\ \midrule
		
	$\hspace{1mm}$ \newline \newline {\bf T-drive}				&	
		 3142, 6834  \newline	 6168, 9513  \newline	 1950, 5896  \newline	 2876, 3260	\newline	 1350, 5970	&
		143, 116	\newline	143, 119	\newline	140, 123	\newline	100, 132	\newline	132, 127	&
		$14.97 \times 16.64$	 \newline	$0.97 \times 0.83$ \newline	$1.16 \times 1.31$ \newline	$0.81 \times 0.51$ \newline	$0.76 \times 0.48$  &   
		10 \newline 10 \newline 10 \newline 10 \newline 10	 \\ \midrule  
	
	$\vspace{2mm}$ \newline {\bf Geolife}				
		&  15, 44 \newline	 15, 125  	\newline  16, 44	\newline  33, 40 	
		&	154, 197  \newline	154, 122	\newline	140, 197	\newline	117, 193
		&		$0.32 \times 0.6$	\newline	$14.37 \times 4.07$ \newline	$0.18 \times 0.32$ \newline	$0.21 \times 0.52$  &
		1 \newline 1 \newline 1 \newline 1 \\  \bottomrule 
\end{tabular}
\end{table}

As a general preprocessing step for all experiments first we remove stationary waypoints (i.e.,  consecutive waypoints with no movement in between) and then remove all trajectories with less than 10 waypoints. The trajectories are randomly split (70/30) into train and test data and we report misclassification error on test data for several classifiers, averaged over 50 random train-test splits.  

The experimental setup for the mistake-driven algorithm is as follows.  We split data into train and test data (70/30) and calculate the threshold value $\eta$ using the training data. Then we apply our method of choosing landmarks (Algorithm \ref{algo: choose landmarks}) 3 times (unless otherwise specified) on training data and opt for one with the lowest training error. This one is then used on test data to report the results in tables (as usual, averaged over 50 test/train splits).

We apply the same methodology for different choices of landmarks $v_Q$, $v_Q^{\exp}$, $v_Q^{\varsigma}$ and using endpoints, when appropriate. 
The generic random landmarks are chosen from normal distribution with mean of all waypoints and standard deviation $4$ times the standard deviation of all waypoints, and are denoted ``Rand $v_Q$''.  The ones that are mistake-driven are denoted ``MD $v_Q$''.  When voting is used it is marked Vote($\cdot$).

\paragraph{Classification methods used.}
Given the various ways of achieving vectorized representations (including just mapping to $\mathbb{R}^4$ via the coordinates of the endpoints), we experiment with Convolutional Neural Networks (CNN in short) and 7 classifiers from $\sklearn$:    
Linear Kernel SVM (LSVM in short), Gaussian Kernel SVM (GSVM in short), Polynomial Kernel SVM (PSVM in short), Decision Tree (DT in short), Random Forest (RF in short), K-Nearest Neighbor (KNN in short) and Logistic Regression (LR in short). Unless otherwise specified the hyperparameters are chosen as follows. In Linear, Gaussian and Polynomial SVM the hyperparameter $C$ is set to 100. The hyperparameters $\gamma$ in Gaussian SVM and $deg$ in Polynomial SVM are set to `$auto$'. Decision Tree is applied with no limit on `$max\_depth$', Random Forest with 50 estimators and KNN with 5 neighbors. All other hyperparameters/parameters are set to the default ones. The parameter $\varsigma$ in $v_Q^{\varsigma}$ vectorization for all experiments in this section is chosen sufficiently large with respect to the length of the rectangular region containing all trajectories in order to reduce the Gaussian weight impact on the vectorization while capturing the orientation. For the CNN we use a 1-layer architecture with 10 (1-dim) convolutional layers in the hidden layer. Padding and stride of 1 are utilized and convolution kernel size is set to 2 or 5 depending on the dimension of the input data.  

We compare these results with KNN (K-nearest neighbor) classifier with $k=5$ estimators from $\sklearn$ library, using 12 different distances: Continuous Fr\'echet  Distance \cite{AG95}, Discrete Fr\'echet  Distance \cite{TEHM1994}, Hausdorff Distance \cite{Hau1915}, DTW (Dynamic Time Warping) \cite{BC94}, fastdtw (an approximation for Dynamic Time Warping) \cite{SC07}, soft-dtw (Soft Dynamic Time Warping) \cite{CB17}, SSPD (Symmetric Segment-Path Distance) \cite{BGLR15}, LCSS (Longest Common Subsequence) \cite{lcss}, EDR (Edit Distance on Real sequence) \cite{edr}, ERP (Edit distance with Real Penalty) \cite{CN04}, $\dQ^{\pi}$ \cite{PT19a} and LSH (computed by binary sketches for Hausdorff distance) \cite{ACKGS2018}. The soft-dtw distance is implemented from GitHub page \url{https://github.com/mblondel/soft-dtw}, fastdtw from PyPI, DTW from $\tslearn$, $\dQ^{\pi}$ from $\mathtt{trjtrypy}$ package by the authors in PyPI, LSH is implemented by the authors, and the rest of the distances use efficient code available from the GitHub page \url{https://github.com/bguillouet/traj-dist}. In all experiments with KNN using LSH, we have applied 20 random circles to get binary sketches utilized in LSH distance. To choose the circles' radius $r$, like computing $\eta$ in the mistake-driven algorithm, we set $r = \|r_1 - r_2\|$, where $r_1$ and $r_2$ are the mean of $(x,y)$-coordinates of trajectories in the $70\%$ of the first and second datasets used as training data. Finally, in all landmark-based experiments in this paper we use 20 landmarks. Indeed, in random $v_Q$, $v_Q^{\exp}$ and $v_Q^{\varsigma}$ we randomly generate 20 landmarks around the curves. For mistake-driven approaches, Algorithm \ref{algo: choose landmarks} chooses 20 landmarks accordingly.  
To make it easy to reproduce or compare against our results, we link to all data sets, and host cleaning and testing methods on a public github page \url{https://github.com/aghababa/Classifying-Spatial-Trajectories}.

All resulting 
figures (Figures \ref{fig: car-bus}, \ref{fig: simulated car-bus}, \ref{fig: Two persons}, \ref{fig: Geolife bar chart}, \ref{fig: Physical car-bus}, \ref{fig: characters}, and \ref{fig: T-drive} and Table \ref{table: trans-modes}) show the average classification error over $50$ trials on random test/train splits.  The error bars show standard-deviation of these $50$ trials.  

\subsection{Car-Bus dataset} \label{sec: car-bus}

\begin{figure}[ht]
\includegraphics[width=0.32 \textwidth]{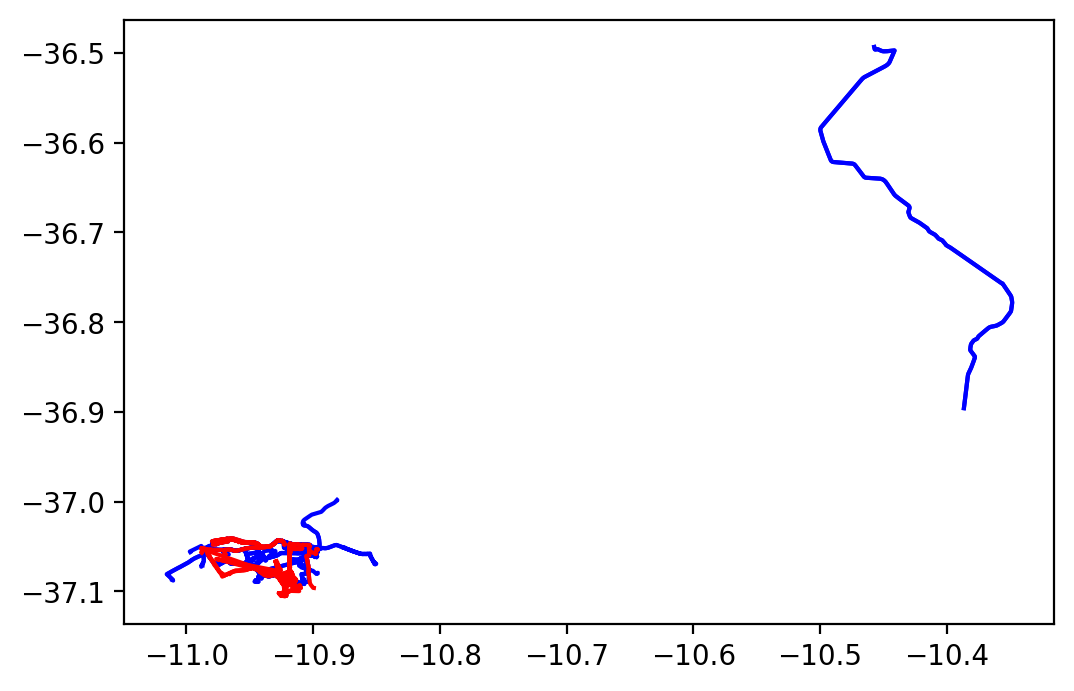}
\includegraphics[width=0.32 \textwidth]{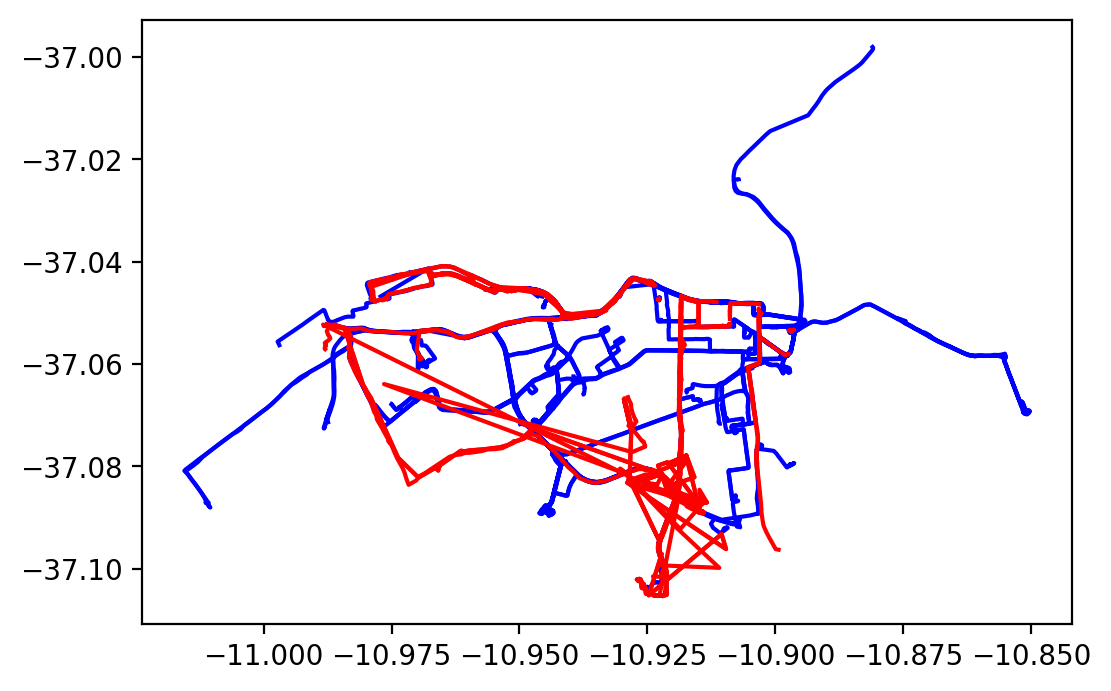}
\includegraphics[width=0.32 \textwidth]{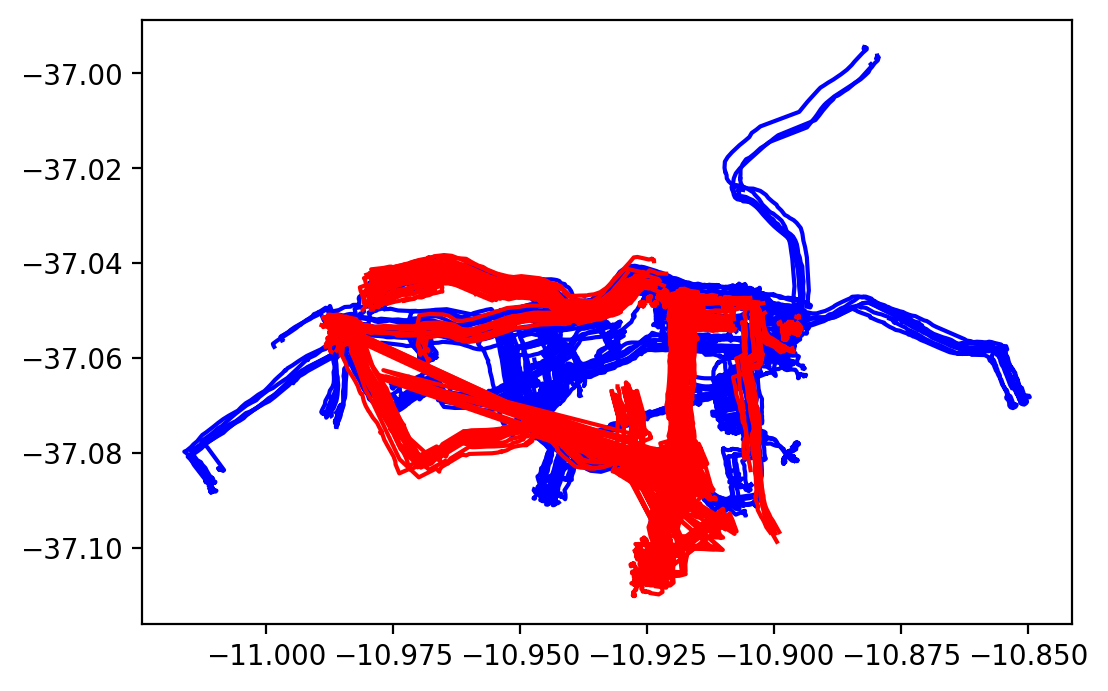}
\caption{Left: Car (blue) and bus (red) trajectories (full original data set used in experiments).  Middle: zooms in for a close up of most interesting data.  Right: shows zoomed in region of the simulated car-bus data set.}  
\label{fig: Simulated-car-bus}
\end{figure}

We first consider the GPS Trajectories Data Set from UCI Machine Learning Repository \cite{GTDS2016}, recorded in Aracuja, Brazil (see Figure \ref{fig: Simulated-car-bus}). There are 87 car and 76 bus trajectories in this dataset. After the preprocessing step described above, a set of 78 car and 45 bus trajectories remained. Moreover, in order to make the problem more challenging, we removed the 2 clear outliers from car and 1 from bus trajectories and ended up with 76 car and 44 bus trajectories. In this experiments $C=10$ is applied for PSVM classifier.

\begin{figure*}[h]
\centering
\includegraphics[width=1 \textwidth]{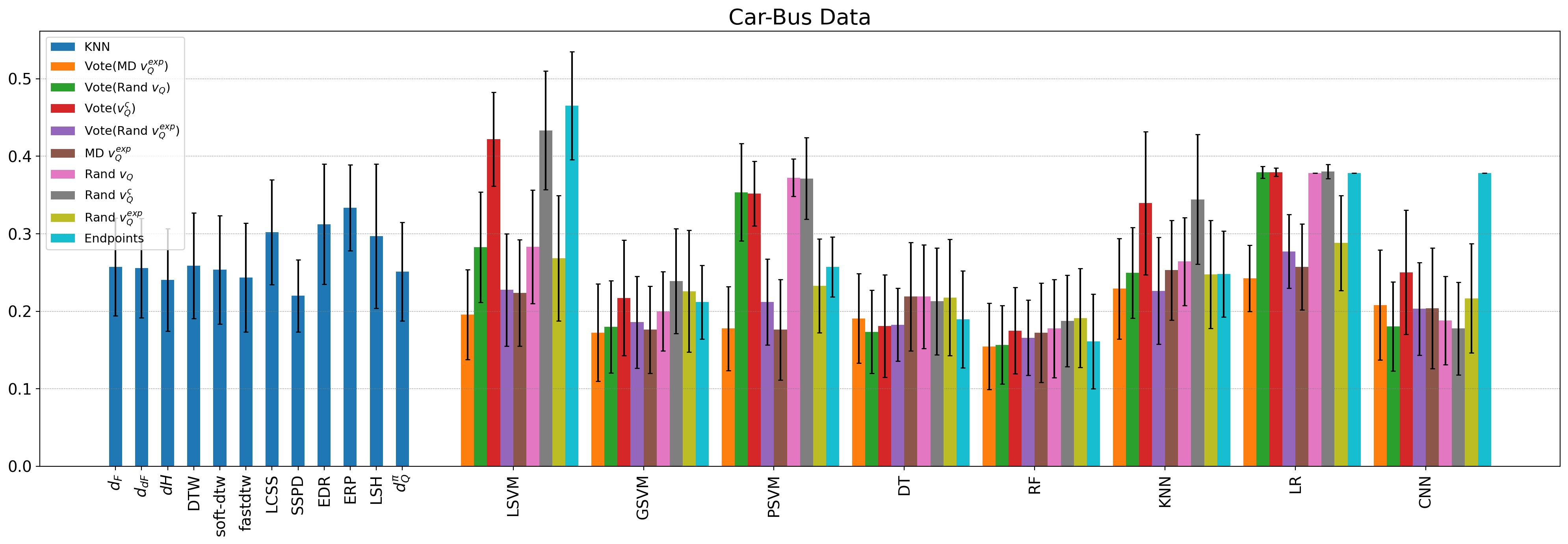}
\caption{Average classification test errors of \underline{Car-Bus} dataset of several classification techniques (in bar charts) with different methods of featurizing the data (in color), mostly based on the way the landmark set $Q$ is chosen. 
Also, average KNN-classification test errors  with various distances.  Parameters: soft-dtw: $\gamma=1e-15$, LCSS and EDR: $\epsilon=0.02$.}
\label{fig: car-bus}
\end{figure*}

The misclassification rates are shown in Figure \ref{fig: car-bus}. The figure shows results of applying a variety of baseline distances, and classification using KNN; all of these approaches have error of at least $22\%$ on the test data. Other methods like decision trees are not generically possible to do with only access to a distance.  
The best misclassification rate on the test error of $15.46\%$ is shown in Figure \ref{fig: car-bus} and achieved by Random Forest using voting technique with mistake-driven landmarks. In general, the mistake-driven landmarks perform better than the similar methods with randomly chosen landmarks or just endpoints as shown in Figure \ref{fig: car-bus}. The $v_Q^{\exp}$ landmarks mostly work a bit better than $v_Q$ and $v_Q^{\varsigma}$ landmarks across techniques. The Random Forest methods do consistently well (with between $15\%$ and $19\%$ error). In addition to the good performance of Random Forest classifier, Gaussian SVM tends to perform well with between $17\%$ and $24\%$ error, but in this case does not achieve the smallest overall test error. In fact, just using endpoints (see Figure \ref{fig: car-bus}), Random Forest achieves the third best result of $16.11\%$ error -- however, other classifier types other than DT, do poorly with this representation, so this does not appear to be a good general purpose way to vectorize the trajectories.

\subsection{Simulated Car-Bus dataset}

In this experiment, to create a larger and more balanced data set, we add some noise to the preprocessed trajectories in car-bus dataset in Section \ref{sec: car-bus} to get 2 noisy copies of each car and 4 noisy copies of each bus trajectories and combine them with the original preprocessed car and bus trajectories. Thus we end up with a roughly balanced data including 226 car and 220 bus trajectories (see Figure \ref{fig: Simulated-car-bus}). To be more precise, we add a noise offset vector $v \in \mathbb{R}^2$ to each waypoint.  We initialize $v = 0.001$, and this is the noise added to the start point. Then before modifying each subsequent waypoint we update $v \leftarrow v + N(0,0.0001)$, which is the two-dimensional normal distribution with mean $(0,0)$ and standard deviation $0.0001$.

\begin{figure*}[h]
\centering
\includegraphics[width=1 \textwidth]{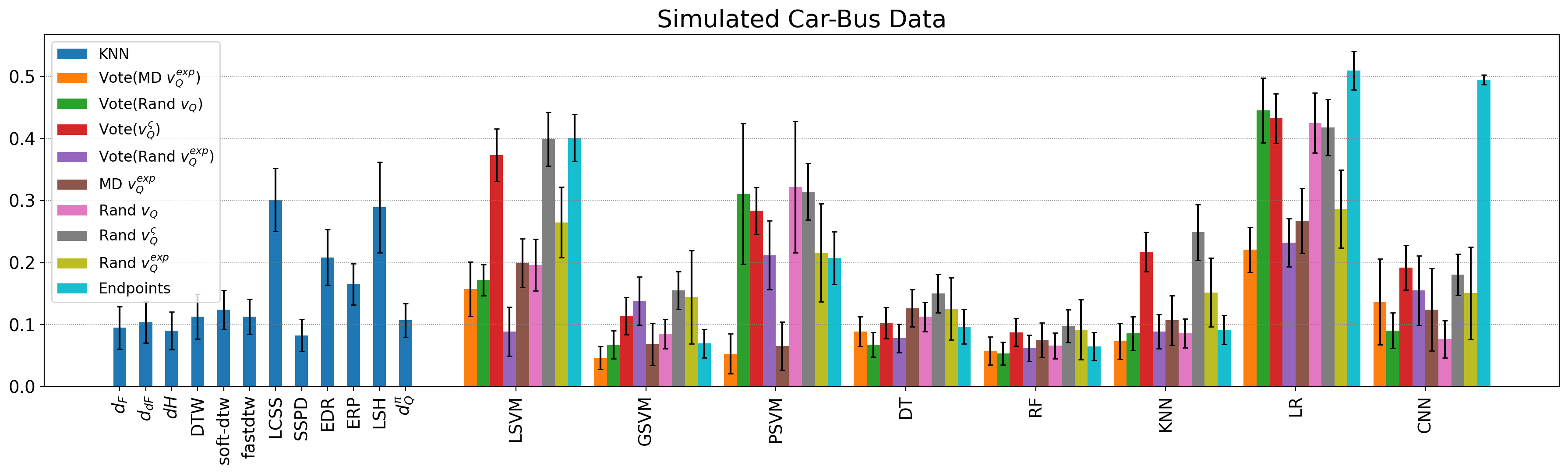}
\caption{Average classification test errors of \underline{Simulated Car-Bus} dataset of several classification techniques (in bar charts) with different methods of featurizing the data (in color), mostly based on the way the landmark set $Q$ is chosen.
Also, average KNN-classification test errors with various distances on trajectories. Parameters: soft-dtw: $\gamma=1e-15$, LCSS and EDR: $\epsilon=0.02$.}
\label{fig: simulated car-bus} 
\end{figure*}

The misclassification rates are presented in Figure \ref{fig: simulated car-bus}. The relative accuracy of classification methods is similar to that of the original car-bus dataset, but the accuracy in this setting is generally better. The left side bars show results of applying 12 baseline distances, and classification using KNN. They all have error of at least $8\%$ on the test data; the best performing one is SSPD-KNN. The overall best misclassification rate of $4.64\%$ shown in Figure \ref{fig: simulated car-bus} is achieved by Gaussian SVM using voting technique of mistake-driven landmarks. In general, the voting method with mistake-driven landmarks does better than other vectorization techniques or just endpoints as shown in Figure \ref{fig: simulated car-bus}. Comparing the performance of featurization methods we observe that random $v_Q$ vectorization technique tends to perform a bit better than $v_Q^{\varsigma}$ and $v_Q^{\exp}$ vectorizations. Perhaps surprisingly, with just using endpoints as the vectorized features (see Figure \ref{fig: simulated car-bus}), Random Forest achieves a low error rate of $6.49\%$; this may be partially explainable due to the relatively low noise of starting points in the noisy copies of each trajectory.   Linear classifiers generally perform poorly on most featurizations.

\subsection{Two persons trajectory data}

\begin{figure}[h]
\centering
\includegraphics[width=0.38 \textwidth]{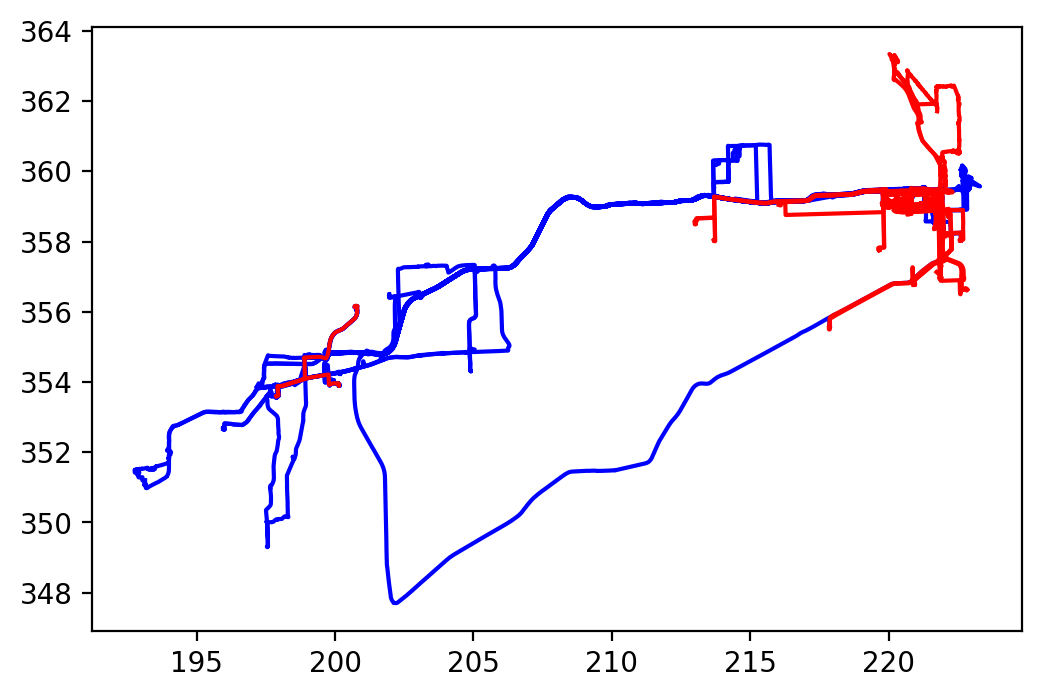}
\caption{GPS trajectories of two members (one in blue and the other in red) of ``Databases and Mobile Computing Laboratory in University of Illinois at Chicago'' during their daily commute for 6 months which are captured in the Cook county and/or the Dupage county of Illinois.}
\label{fig: Two-persons-data}
\end{figure}

The trajectory data in this experiment was obtained from the GPS carried by two members of ``Databases and Mobile Computing Laboratory in University of Illinois at Chicago'' during their daily commute for 6 months (see Figure \ref{fig: Two-persons-data}). The dataset was created in 2006 and is available for public at \url{https://www.cs.uic.edu/~boxu/mp2p/gps_data.html}. Each trajectory represents a continuous trip of a member in Cook county and/or Dupage county, Illinois. One of the persons has 124 trajectories and the other 89 trajectories. We removed all stationary waypoints and trajectories with less than 10 waypoints as the general preprocessing step; however, the number of trajectories for each person did not change. Similar to other experiments the goal is to classify each person's trajectory.

Most of the trajectories in this dataset are very long and thus the running time of KNN classifier with almost all distances are high. Specially, the continuous Fr\'echet distance did not complete in 48 hours, and so we do not report any results. Note that since our vectorization methods are efficient using optimized Euclidean libraries, they are very fast in comparison with KNN classifier using variety of distances (running times are discussed more in Section \ref{Discussion}).

\begin{figure*}[h]
\centering
\includegraphics[width=1 \textwidth]{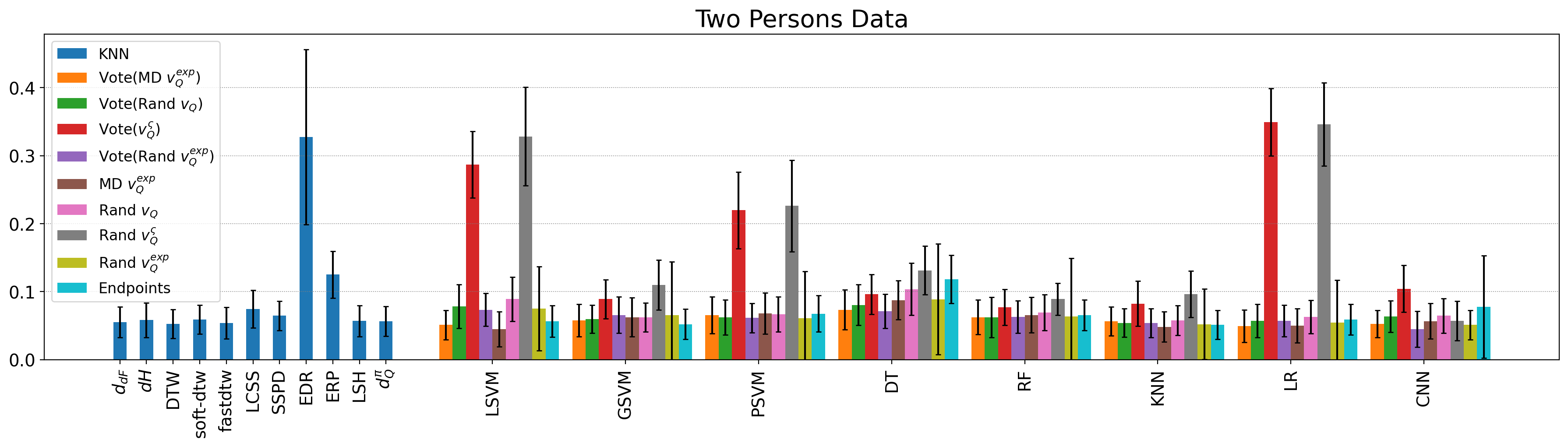}
\caption{Average classification test errors of \underline{Two Persons} dataset of several classification techniques (in bar charts) with different methods of featurizing the data (in color), mostly based on the way the landmark set $Q$ is chosen. Also, average KNN-classification test errors with various distances. Parameters: soft-dtw: $\gamma=1e-10$, LCSS and EDR: $\epsilon=0.1$.}
\label{fig: Two persons}
\end{figure*}

The results in Figure \ref{fig: Two persons} show the misclassification errors on test data. The lowest misclassification rate of $4.49\%$ is achieved by mistake-driven landmarks using Linear SVM and also by Convolutional neural networks applying voting technique on $v_Q^{exp}$-vectorization. Other classifiers included cannot achieve an error rate less than $4.83\%$ (which is also gained by mistake-driven landmarks) and KNN with variety of popular distances cannot do better than $5.26\%$. All other vectorization techniques  tend to work roughly similar on this dataset, where all could get under $6\%$ misclassification rate with at least one classifier. It seems that the direction of trajectories is not essential as the $v_Q^{\varsigma}$-vectorization did comparatively poorly with about $8\%$ test error applying Random Forest and KNN classifier using voting technique except CNN that could get $5.69\%$ misclassification rate.

\subsection{Characters dataset}

\begin{figure}[h] 
\centering
\includegraphics[width=0.19 \textwidth]{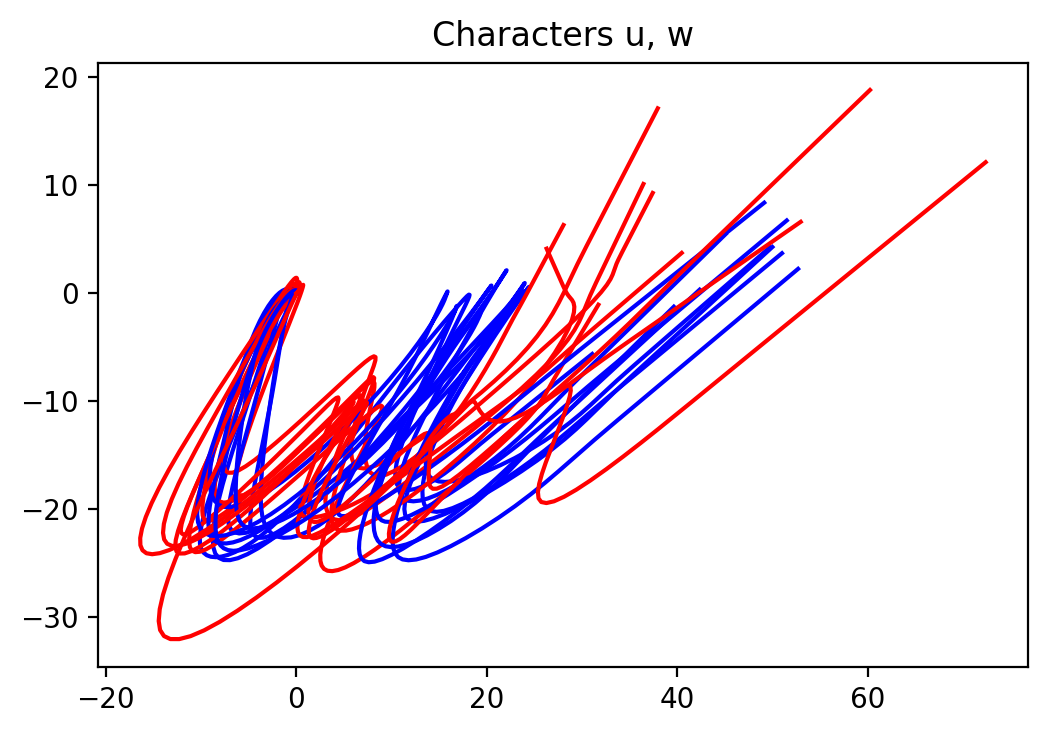}
\includegraphics[width=0.19 \textwidth]{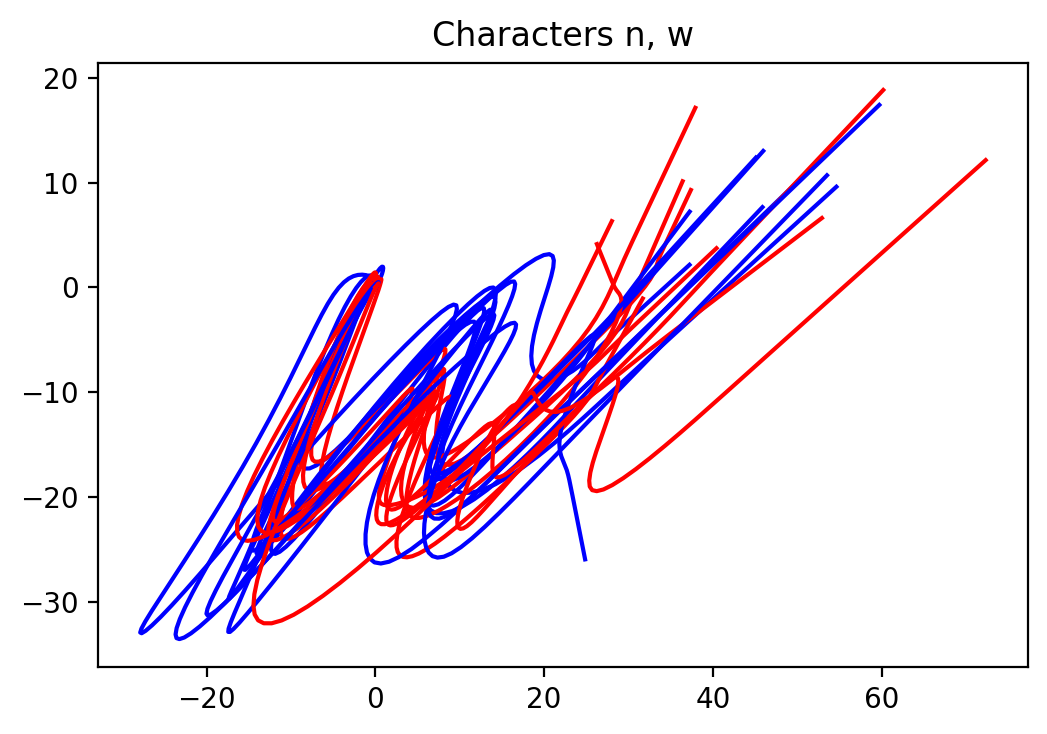} 
\includegraphics[width=0.19 \textwidth]{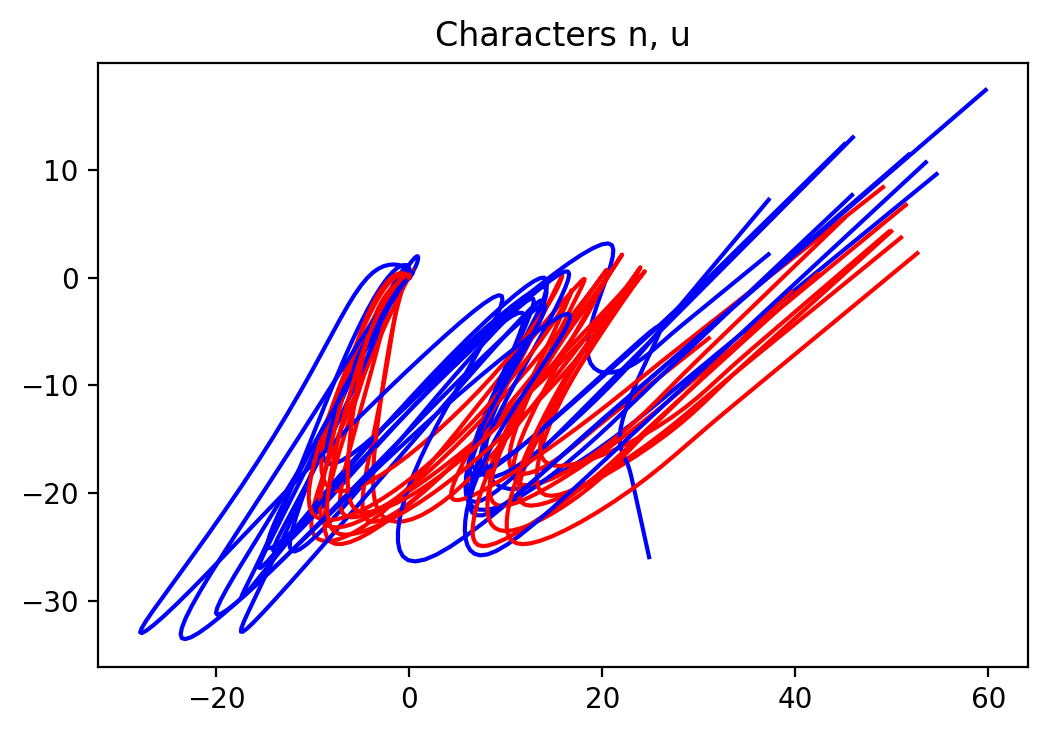} 
\includegraphics[width=0.19 \textwidth]{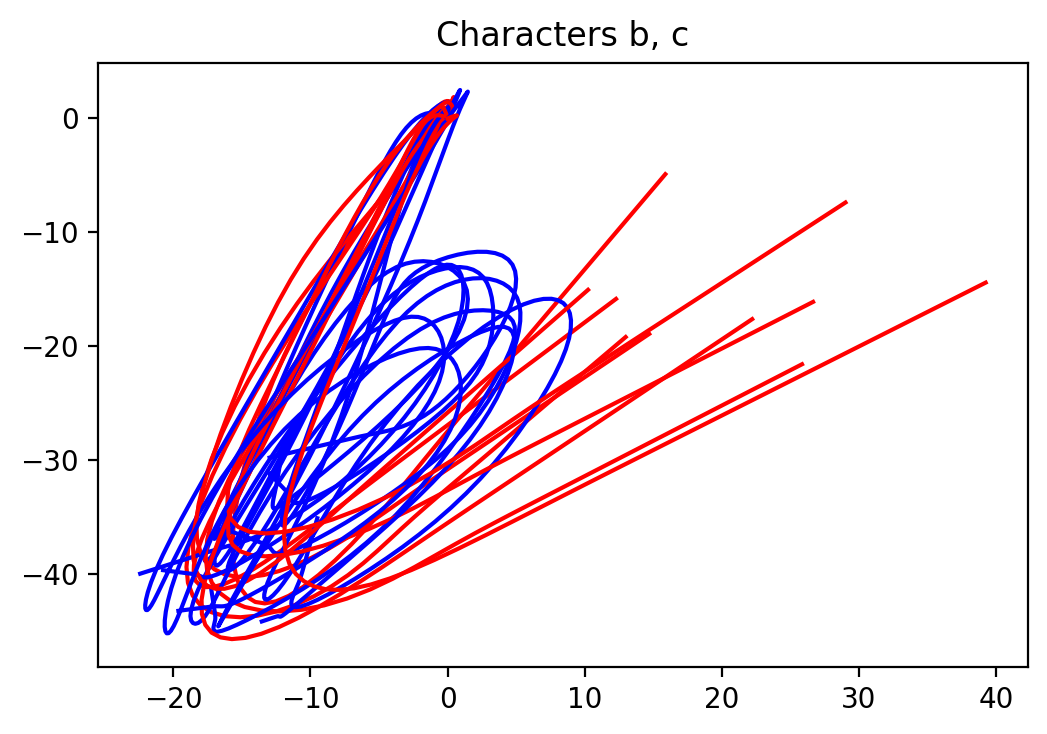} 
\includegraphics[width=0.19 \textwidth]{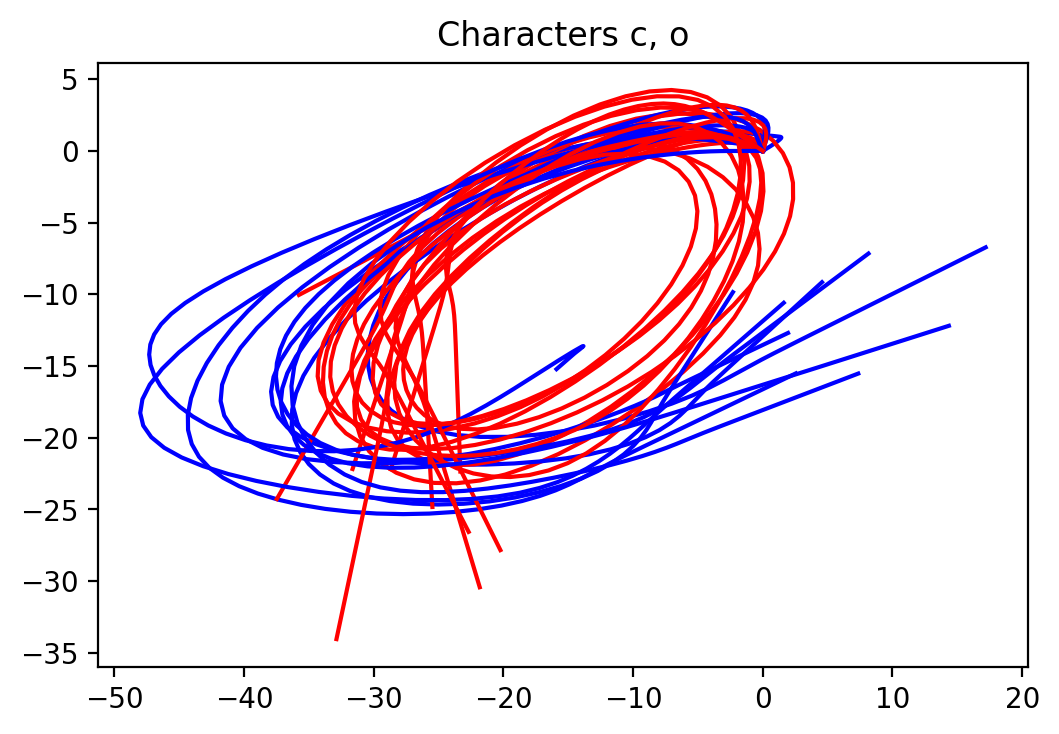} 
\caption{A subset of trajectories (10 trajectories of each character) are presented.}
\label{fig: Characters-data}
\end{figure}

For a change of pace, this experiment is done on the Character Trajectories Data Set from UCI Machine Learning Repository that consists of handwritten characters captured using a WACOM tablet -- so is not mobility data.  We chose 5 similar pairs of letters $\{$(u, w), (n, w), (n, u), (b, c), (c, o)$\}$ (see Figure \ref{fig: Characters-data}) to perform a binary classification for each pair. For other pairs of characters we did a binary classification with $v_Q$-vectorization using random landmarks and got near zero test error. The selected pairs are the more challenging ones. Here $\varsigma=100$ is considered for $v_Q^{\varsigma}$-featurization.

\begin{figure*}[h]
\centering
\includegraphics[width=1 \textwidth]{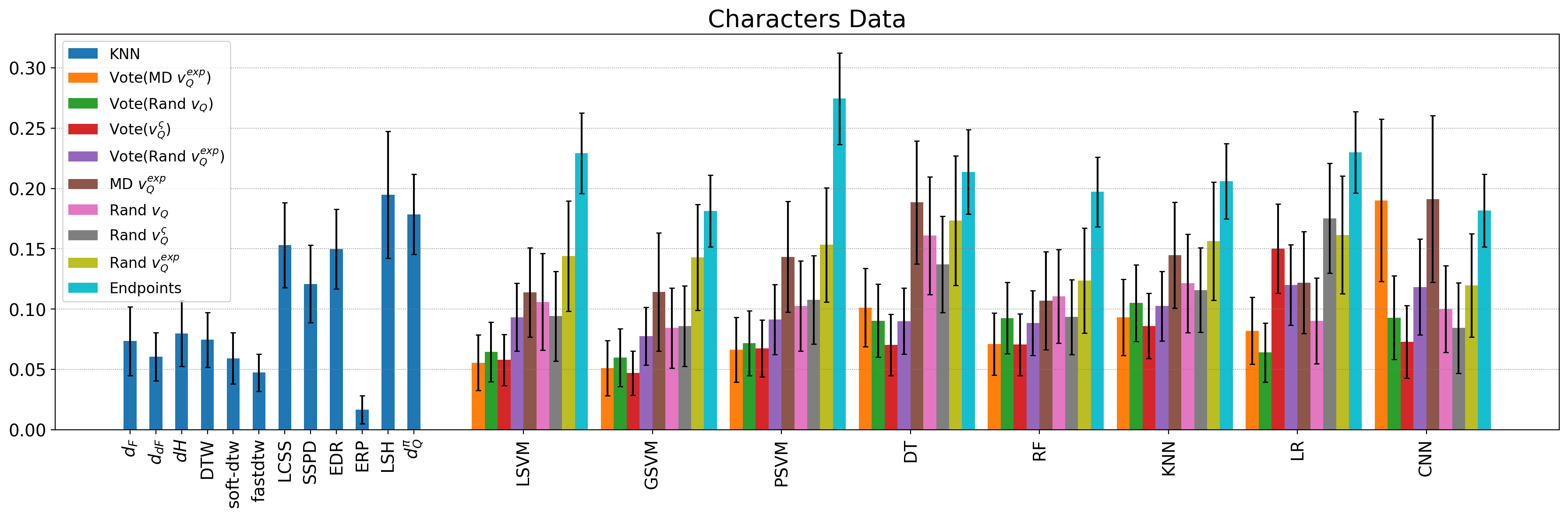}
\caption{Average classification test errors of 5 pairs from \underline{Characters} dataset of several classification techniques (in bar charts) with different methods of featurizing the data (in color), mostly based on the way the landmark set $Q$ is chosen.
Also, average KNN-classification test errors  with various distances. Parameters: soft-dtw: $\gamma=0.15$, LCSS: $\epsilon=1$ and EDR: $\epsilon=1,2$.}
\label{fig: characters}
\end{figure*}

Aggregated test errors for 5 chosen pairs are given in Figure \ref{fig: characters}. The KNN classifier using edit distance with real penalty (ERP) achieves the best misclassification rate. There is a considerable gap between the best misclassification rate of $1.67\%$ and the next best performed classifier with $4.70\%$ misclassification rate which is achieved by Gaussian SVM classifier employing $v_Q^{\varsigma}$-vectorization with voting technique.  
KNN with fastdtw also generated a low error rate of $4.73\%$.  Other vectorized GSVM classifiers did well, but most others had misclassification rate higher than $6\%$, and often above $10\%$. 
Generically, however, the mistake-driven vectorized technique, as well as Vote($v_Q^{\varsigma}$) and endpoints, tends to work better than other landmark-based featurization methods across all classifiers, which are better than other KNN-based ones. The best classifiers among these are SVM-based ones, Random Forest, and Fr\'echet- or DTW-based KNN classifiers.

\subsection{T-drive Trajectory data set}

\begin{figure}[h] 
\centering
\includegraphics[width=0.19 \textwidth]{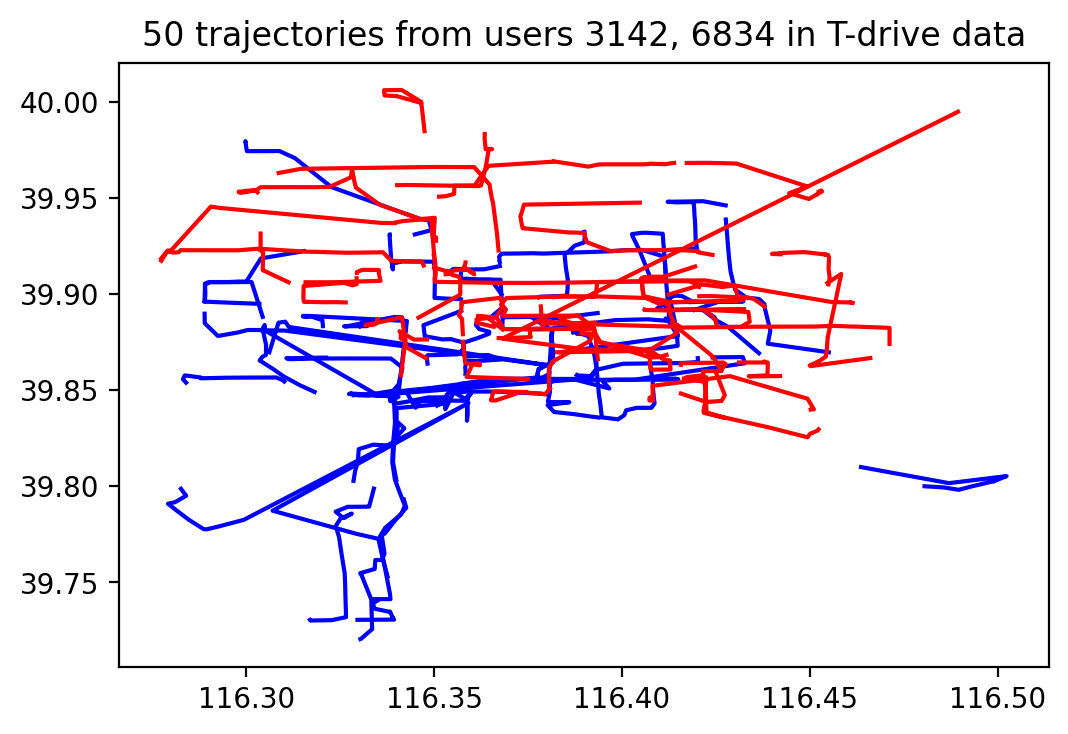}
\includegraphics[width=0.19 \textwidth]{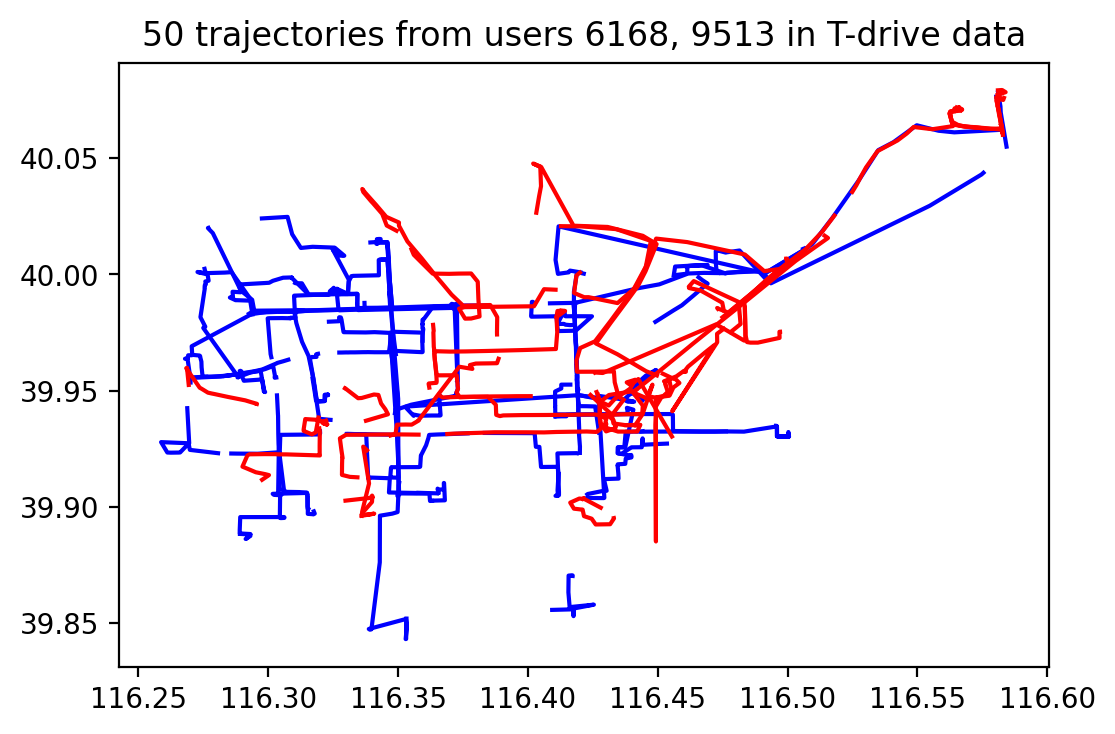} 
\includegraphics[width=0.19 \textwidth]{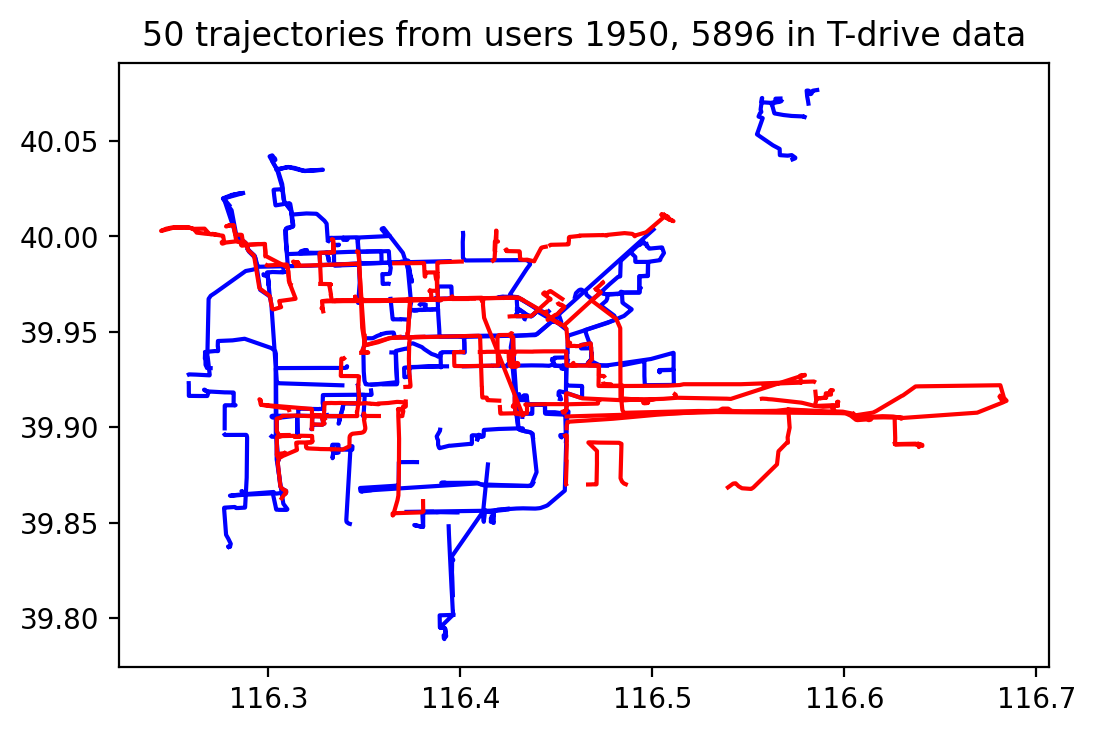} 
\includegraphics[width=0.19 \textwidth]{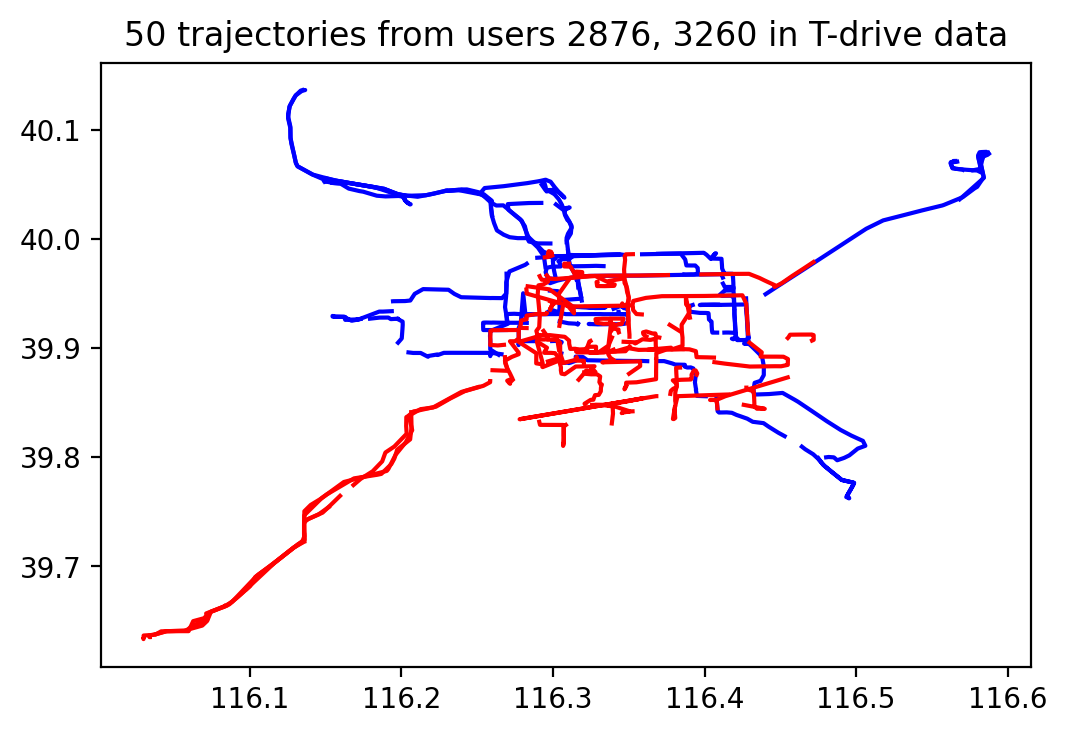} 
\includegraphics[width=0.19 \textwidth]{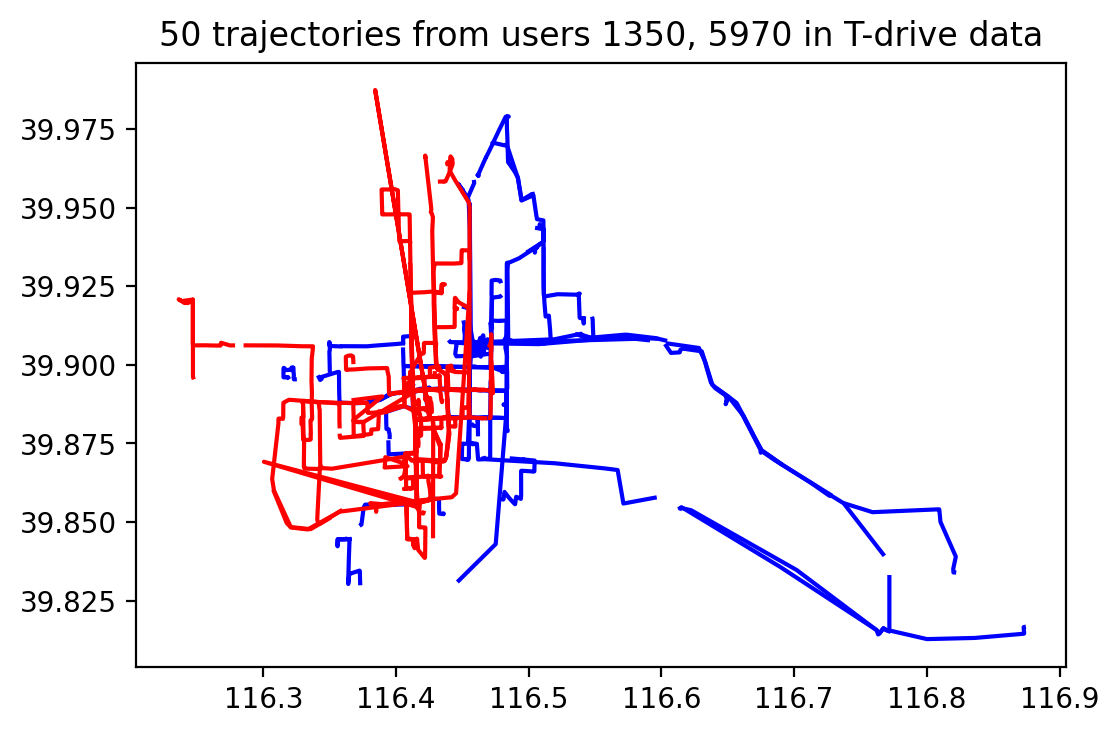} 
\caption{A subset of trajectories (50 trajectories of each user) of each pair of taxies from T-drive dataset are presented.}
\label{fig:T-drive-data} \vspace{-1mm}
\end{figure}

In this experiment we consider the T-drive Trajectory data set released by Microsoft in 2011 \cite{YZZXXSH2010,YZXSH2011}. It consists of trajectories of 10,357 taxis captured by GPS within 6 months in Beijing. Each taxi's trajectories are stacked together so that each taxi has one very long trajectory. We have chosen 5 pairs of taxis ((3142, 6834), (6168, 9513), (1950, 5896), (2876, 3260), (1350, 5970)) with high misclassification rate (measured with both KNN classifier with different distances and featurization via $v_Q$ with 20 random landmarks). Then, after removing stationary waypoints, we partitioned each selected taxi's long trajectory to short trajectories with trip duration of at most 20 minutes and removed all trips with less than 10 and more than 200 waypoints. This way we ended up with between 100 and 200 trajectories for each taxi (see Table \ref{table: data sets} for exact numbers and see Figure \ref{fig:T-drive-data}). \vspace{-1mm}

\begin{figure*}[h]
\centering
\includegraphics[width=0.96 \textwidth]{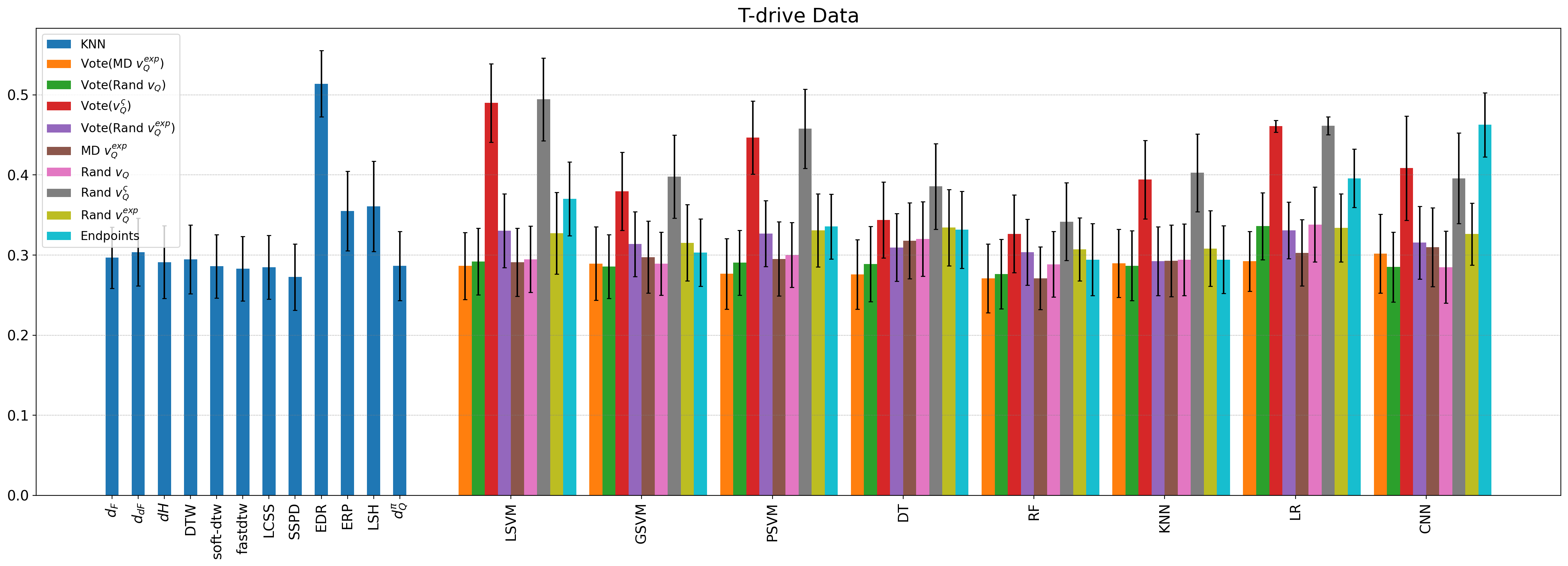}
\caption{Average classification test errors of 5 pairs from \underline{T-drive Trajectory} dataset of several classification techniques (in bar charts) with different methods of featurizing the data (in color), mostly based on the way the landmark set $Q$ is chosen.
Also, average KNN-classification test errors  with various distances. Parameters: soft-dtw: $\gamma=1e-14$, LCSS: $\epsilon=0.0055$ and EDR: $\epsilon=0.005$.}
\label{fig: T-drive}
\end{figure*}

The misclassification rates on test data are reported in Figure \ref{fig: T-drive}. The best performing classifier is Random Forest with $27.1\%$ error rate applying with/without voting technique with the mistake-driven landmarks. KNN with SSPD is the next well performed classifier with error rate of $27.26\%$. KNN with 4 distances, the mistake-driven algorithm with all classifiers but CNN, and voting technique with $v_Q$-vectorization with all classifiers but LR could get misclassification rates between $28\%$ and $29\%$. We can easily observe that the mistake-driven technique tends to do a better job. Furthermore, in contrast to Characters dataset, we see that KNN classifier with ERP could not achieve a low test error.

\subsection{Geolife GPS Trajectory data set} \label{sec: Geolife}

\begin{figure}[h] 
\centering
\includegraphics[width=0.24 \textwidth]{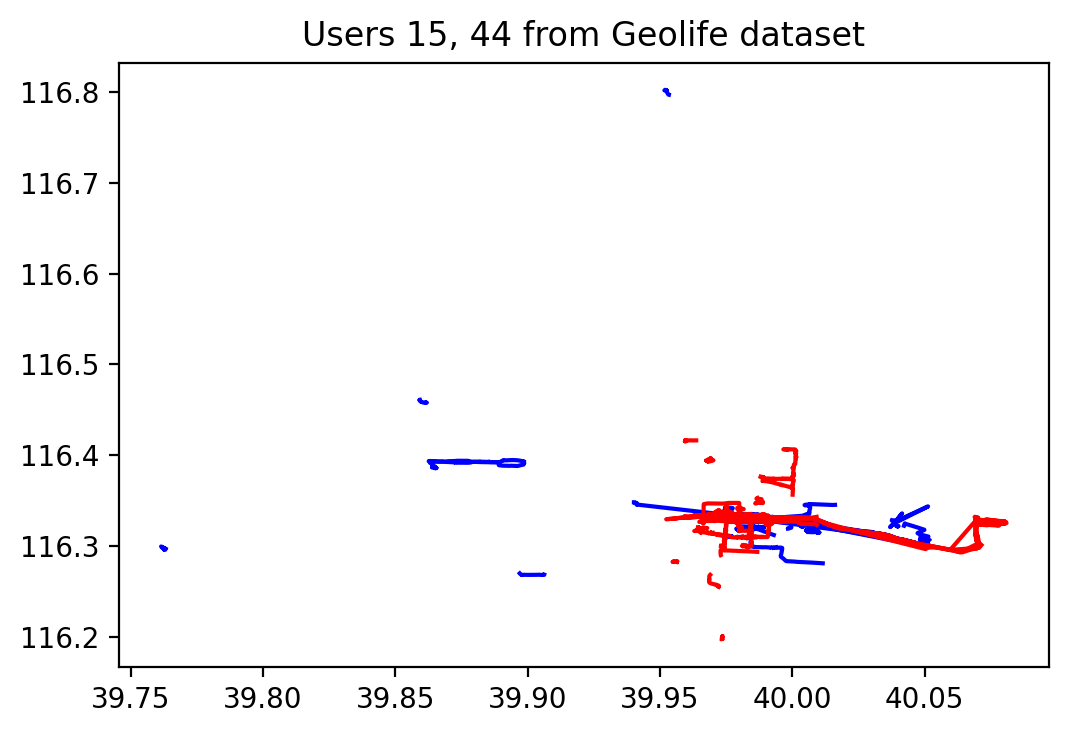}
\includegraphics[width=0.24 \textwidth]{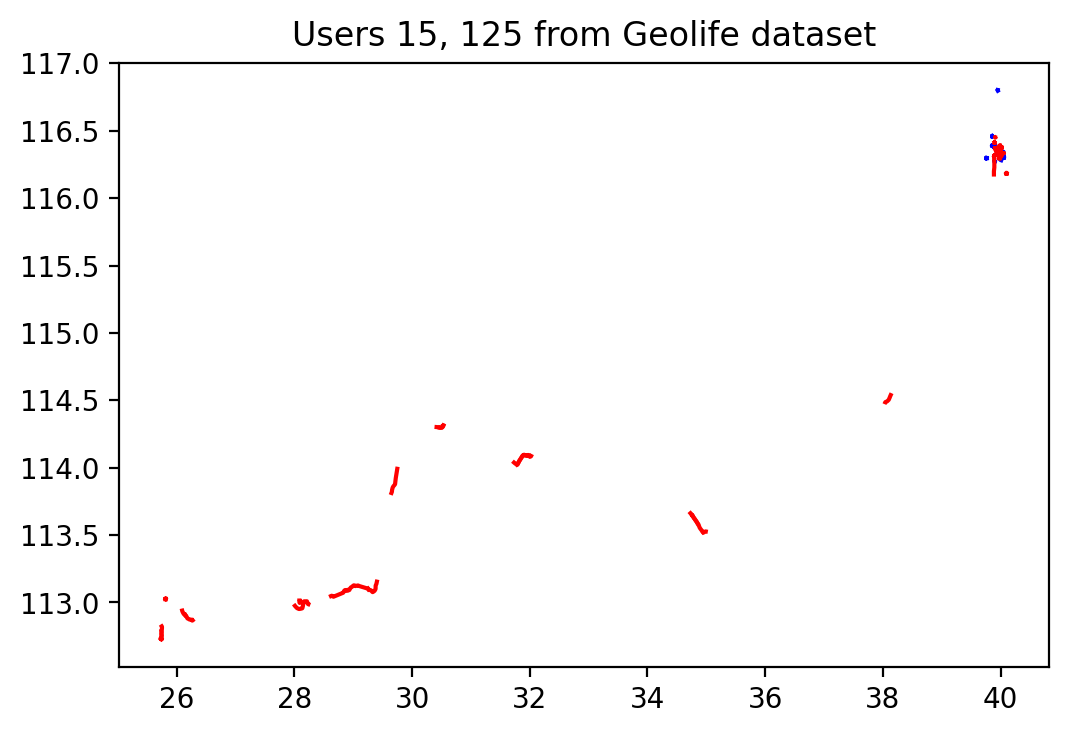} 
\includegraphics[width=0.24 \textwidth]{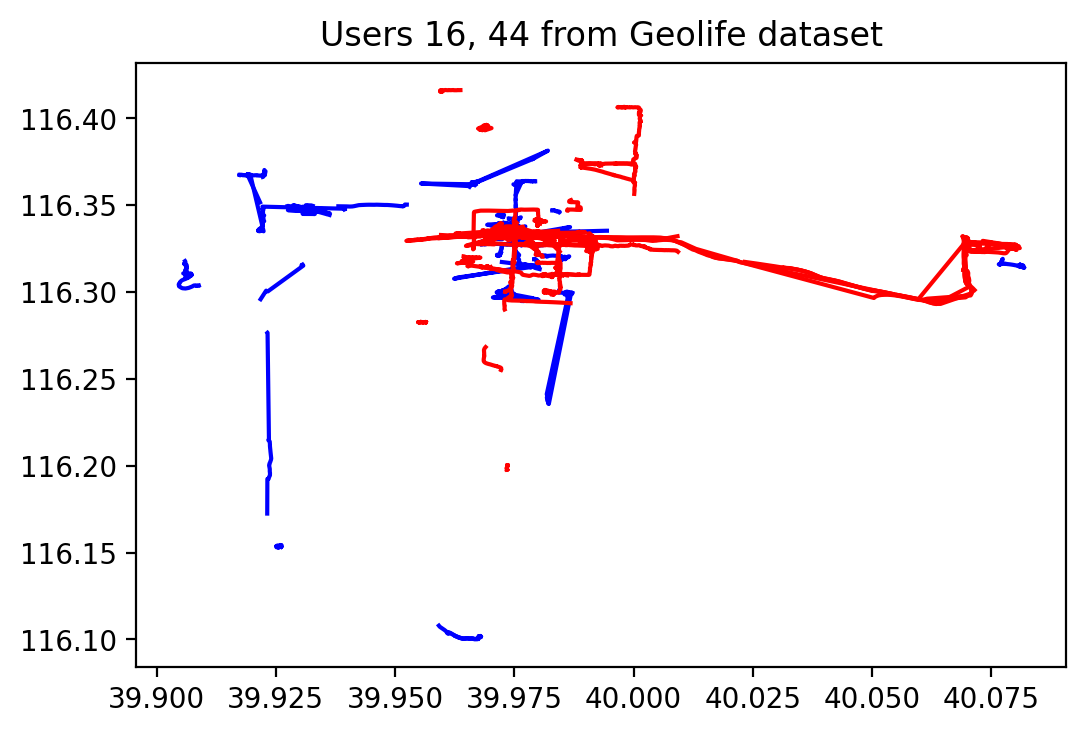} 
\includegraphics[width=0.24 \textwidth]{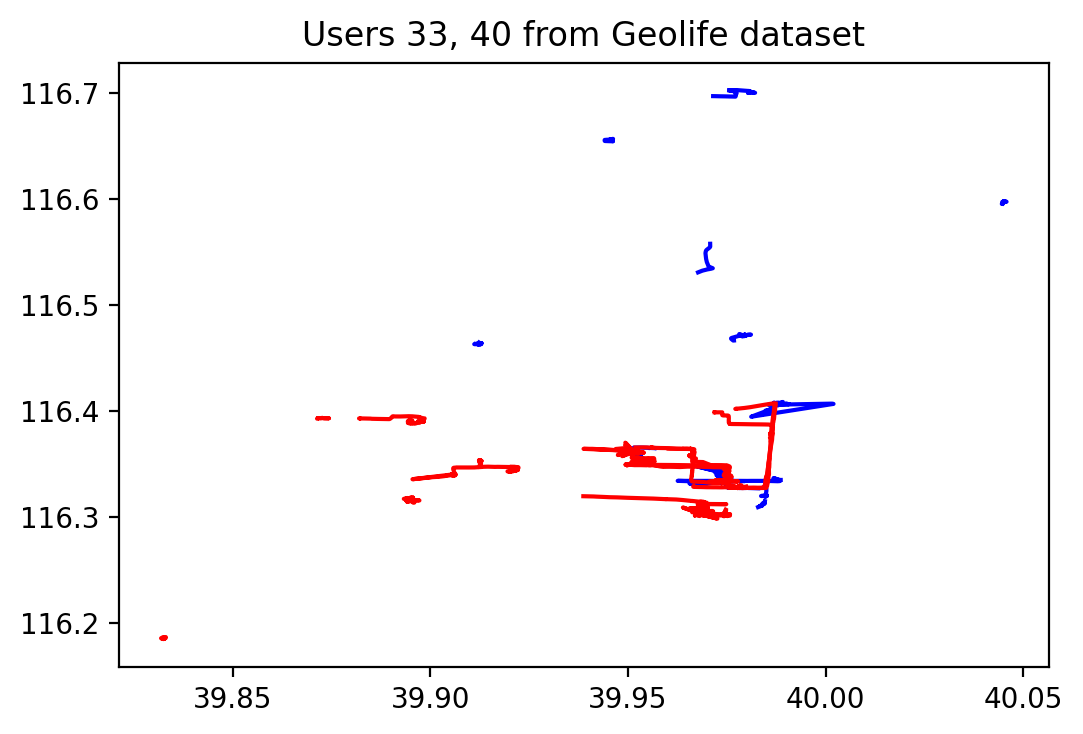} 
\caption{4 pairs of users from Geolife trajectory data after applying 20 minutes partitioning.}
\label{fig:Geolife-data} 
\end{figure}

The dataset we employ here is Geolife GPS Trajectory dataset \cite{geolife-gps-trajectory-dataset-user-guide}, which was released by Microsoft in 2012. It consists of trajectories of 182 users from 2007 to 2012. In total there are 17,621 trajectories which are mostly recorded in Beijing, China. Here we took 4 pairs of users $\{$(15, 44), (15, 125), (16, 44), (33, 40)$\}$ with a high binary misclassification error rate using both KNN (with different distances) and $v_Q$ featurization via a random set of landmarks. Then we tried to classify trajectories of users in each pair from each other. In the preprocessing phase after removing stationary waypoints, like in T-drive dataset, we partitioned each trajectory to trips with a duration of at most 20 minutes, and then applied our standard filtering and cleaning. We ended up with users including between 100 and 200 trajectories (see Figure \ref{fig:Geolife-data}).

Most raw trajectories were quite long and so classification with KNN using most distances is inefficient as their run time is quadratic in the number of waypoints of trajectories. Therefore, we used the 20 minutes threshold to partition trajectories into smaller and perhaps meaningful (sub-)trajectories. The landmark-based vectorization methods' complexities are linear in the number of waypoints of trajectories and thus very efficient.

\begin{figure*}[h]
\centering
\includegraphics[width=1 \textwidth]{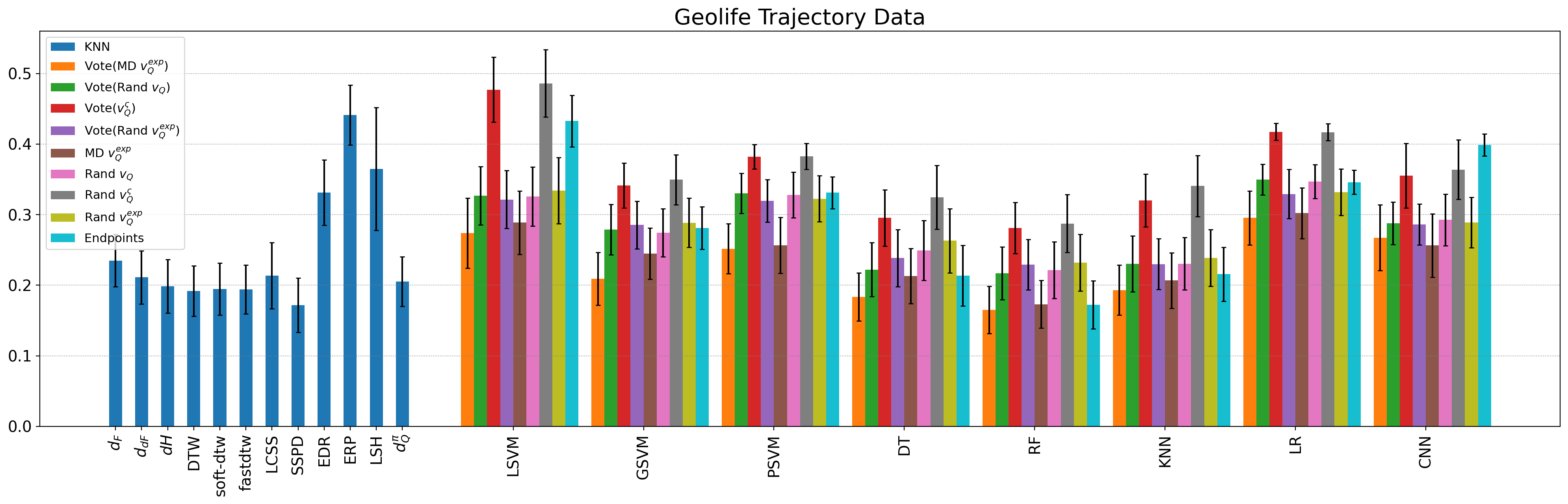}
\caption{Average classification test errors of 4 pairs from \underline{Geolife GPS Trajectory} dataset of several classification techniques (in bar charts) with different methods of featurizing the data (in color), mostly based on the way the landmark set $Q$ is chosen.
Also average KNN-classification test errors with various distances on trajectories. Parameters: soft-dtw: $\gamma=1e-15$, LCSS and EDR: $\epsilon=0.001$.}
\label{fig: Geolife bar chart}
\end{figure*}

The aggregated misclassification results for the chosen 4 pairs are given in Figure \ref{fig: Geolife bar chart}. The best performed method is the mistake-driven algorithm with voting technique using Random Forest with $16.52\%$ error rate on test data. The KNN classifier with SSPD, endpoint classification with Random Forest and the mistake-driven landmarks with Random Forest gained next lowest misclassification rates between $17\%$ and $18\%$. In general, Random Forest tends to perform better than other classifiers across all vectorization methods. Moreover, like in T-drive dataset and unlike Characters dataset, on the Geolife trajectory dataset KNN-ERP performed poorly with a high misclassification rate of $44.13\%$ on test data.

\section{Including Other Features} \label{sec: other features}

Recall that a trajectory is a finite sequence of waypoints, which we transform into a curve connecting consecutive waypoints by line segments.  In this section, we associate a time value $t_i$ with each waypoint as well -- this is common but not universal in their collection.  In addition to an {\it average length} of a segment, this allows us to define other features, namely {\it average velocity}, {\it acceleration} and {\it jerk}.  Using these features, in addition to $v_Q$ and $v_Q^{\varsigma}$, we can have the following feature mappings which are a combination of $v_Q$ or $v_Q^{\varsigma}$ and average length, velocity, acceleration and jerk. So these feature mappings will be capable of capturing both geometric and physical aspects of trajectories. 
\[
v_Q^+(\gamma) = (v_Q(\gamma), \text{length}, \text{velocity}, \text{acceleration}, \text{jerk}) \in \mathbb{R}^{\mid Q \mid +4},
\]
\[
v_Q^{\varsigma +}(\gamma) = (v_Q^{\varsigma}(\gamma), \text{length}, \text{velocity}, \text{acceleration}, \text{jerk}) \in \mathbb{R}^{\mid Q \mid +4}.
\]
To be able to compare the contribution of geometric and physical attributes, we will evaluate misclassification rates with geometric features and physical features separately as well.

\subsection{Car-Bus}

\begin{figure}[h]
\centering
\includegraphics[width=0.6 \textwidth]{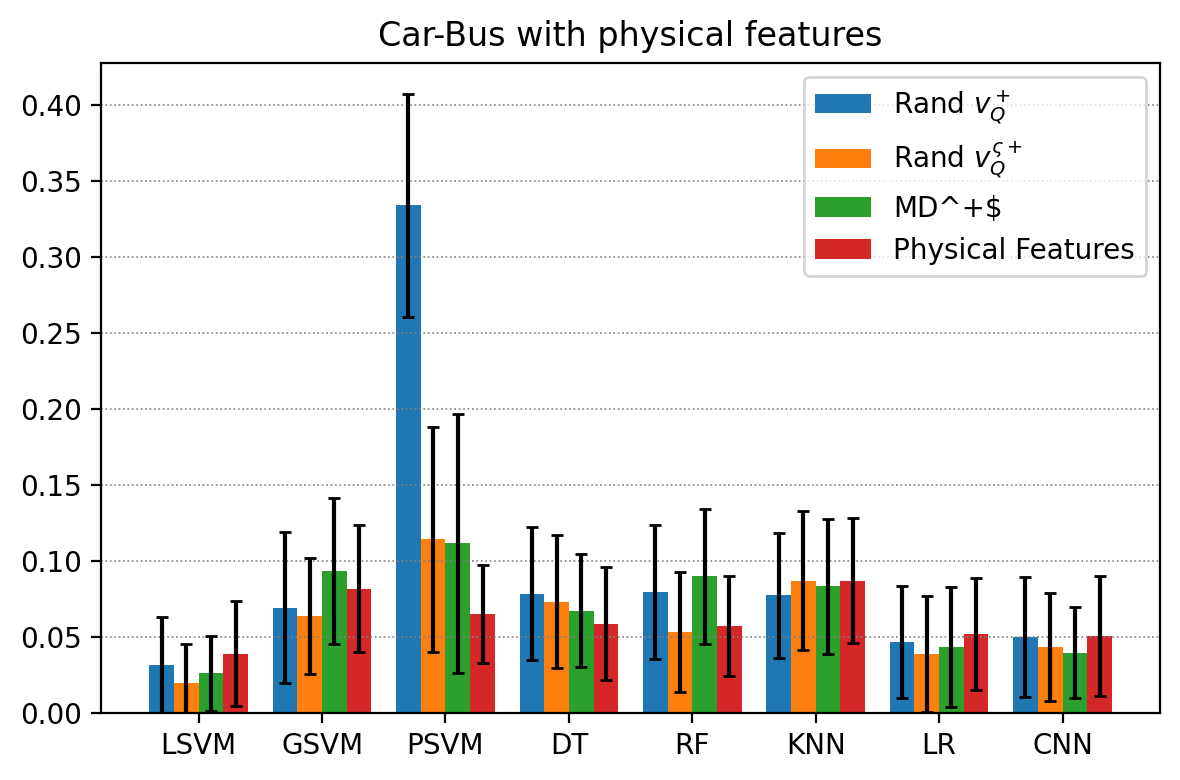}
\caption{Car-Bus classification test errors with different classifiers and vectorizations}
\label{fig: Physical car-bus}
\end{figure}

We reconsider the car-bus dataset utilized in Section \ref{sec: car-bus}. We use the same preprocessing step here and try to classify car versus bus trajectories using 20 landmarks. As it can be observed from Figure \ref{fig: Physical car-bus}, the lowest misclassification rate of $1.95\%$ is achieved by Linear SVM using $v_Q^{\varsigma +}$ featurization (i.e., the signed feature mapping $v_Q^{\varsigma}$ plus physical features). The next well performed method is again Linear SVM but with mistake-driven algorithm with $2.59\%$ test error, where we combined the physical features with geometric features from landmarks. Recall that the best performance without physical features in Figure \ref{fig: car-bus} was $15.46\%$ misclassification rate; so this is a significant improvement. Also an improvement over $3.89\%$, which is the best error rate with \emph{only} physical features, which uses Linear SVM again.  
Looking at linear classifiers (Linear SVM and Logistic Regression) in Figure \ref{fig: Physical car-bus} and misclassification rates reported in Figure \ref{fig: car-bus}, one may conclude that adding physical features makes the car-bus data almost linearly separable. Other classifiers with all kinds of vectorizations have at least $5\%$ error rate on test data. Figure \ref{fig: Physical car-bus} shows that on car-bus dataset the $v_Q^{\varsigma +}$ vectorization generically works better than other vectorization techniques in presence of physical features. $C=100$ (LSVM and GSVM), $\gamma=auto$ (GSVM), $C=1000$ and ${\rm deg}=auto$ (PSVM), ${\rm n\_estimators}=50$ (RF) and ${\rm n\_neighbors}=5$ (KNN) are selected hyperparameters.

\subsection{Transportation Modes}

The Geolife GPS Trajectory dataset used in Section \ref{sec: Geolife} has a standard mode-of-transportation prediction task. Among 182 users 69 of them have labeled their trajectories with transportation modes such as walk, bike, bus, car, taxi, subway, railway, train, airplane, motorcycle, run. As other modes are in extreme minority in the labeled data, we only consider walk, bike, bus, car, taxi, subway, railway, and train. Moreover, as it is recommended in the user guide of data, we regard car and taxi as one transportation mode, namely car, and similarly, we regard all three labels train, railway and subway as train. Therefore, we deal with 5 labels (walk, bike, bus, car, and train).

In this experiment we try to identify transportation modes of GPS trajectories in the data, so it is a multi-class classification task. First we apply our usual preprocessing step including removing stationary points and trajectories less than 10 waypoints. Since we would like to fairly compare our results with the results from state-of-the-art papers, we try to apply the experimental setup used in \cite{DCHR2020, DH2018}, for example. Thus, in this experiment we use 80-20 train-test split. Moreover, each trajectory is partitioned into short trajectories if the time interval between two consecutive waypoints exceeds the 20 minutes threshold according to the recommendation in \cite{ZX2008}, which is also employed in the aforementioned references. The distribution of transportation modes after the preprocessing step is as follows: walk: 2060, bike 1110, bus: 1094, car: 975, train: 359, which in total are 5174 trajectories.

The misclassification rates on test data are provided in Table \ref{table: trans-modes}. The hyperparameter $\varsigma$, like in Section \ref{sec: Geolife} is set to $1$. As it can be viewed from Table \ref{table: trans-modes}, Random Forest with $v_Q^+$ vectorization achieves the best misclassification rate of $11.89\%$. Furthermore, Random Forest consistently performs better than other classifiers on all vectorization methods, where the next lowest test errors rates of $15.46\%$ and $16.42\%$ are obtained by Random Forest with MD$^+$ and $v_Q^{\varsigma +}$ featurizations. As it can be seen, the mistake-driven technique for choosing landmarks does not work as well as random choice of them. We believe this is because (a) the mistake-driven algorithm is designed for binary classification tasks but this task is a multi-class classification task, and (b) the mistake-driven algorithm helps optimize the landmarks, but does not factor in the physical characteristics which are critical for this task. Moreover, without using physical features, with random choice of landmarks we could achieve misclassification error rate of $18.08\%$ using Random Forest whilst using only physical features the best gained test error rate is $22.73\%$. Therefore, the contribution of both physical and geometric features helps to improve the misclassification rate.

\begin{table}[h]
\caption{Transportation modes classification test errors of Labeled Geolife Trajectory  dataset with different classifiers and different vectorizations. Hyperparameters: LSVM: $C=1e10$, GSVM: $C=\gamma=1000$, PSVM: $C=1000$, ${\rm deg} =3$, RF: ${\rm n\_estimators}=100$ and KNN: ${\rm n\_neighbors}=5$.} 
\label{table: trans-modes} 
\centering
\begin{tabular}{rcccc}
\toprule
    {\bf FM}	&	Rand $v_Q^+$		    &	Rand $v_Q^{\varsigma +}$  	&	MD$^+$ 	\\ \midrule
    {\bf Clf}  	&	 {\bf Mean} ({\bf Std}) &	{\bf Mean} ( {\bf Std})	    &	{\bf Mean} ( {\bf Std})   \\ \midrule
    LSVM  		&	0.2869		(0.0139)	&	0.4059	(0.0138)		    &   0.4416	(0.0178)	\\ 
    GSVM  	    &	0.2068		(0.0126)	&	0.5696	(0.0145)		    &	0.2251	(0.0160)	\\ 
    PSVM  	    &	0.6335		(0.0326)	&	0.5379	(0.0149)		    &	0.6292  (0.0114)	\\ 
    DT   	    &	0.1973		(0.0116)	&	0.2688	(0.0123)		    &	0.2289	(0.0131)	 \\ 
    RF  	    & {\bf 0.1189}  (0.0094)	&	0.1642	(0.0100) 		    &	0.1546	(0.0085)	\\ 
    KNN   		&	0.2086		(0.0105)	&	0.2611	(0.0111) 		    &	0.2303	(0.0300)	\\ 
    LR          &	0.6564		(0.1775)	&	0.6840	(0.1552) 		    &	0.6261	(0.0109) \\  \vspace{-3mm}
\end{tabular} 
\begin{tabular}{rccc}
\toprule
    {\bf FM}	&	    Rand $v_Q$		    &	Rand $v_Q^{\varsigma}$	&	Physical Features	    \\ \midrule
    {\bf Clf}   &	{\bf Mean} ({\bf Std})	&	 {\bf Mean} ({\bf Std}) &	{\bf Mean} ({\bf Std})	 \\ \midrule 
        LSVM  	&	0.4959	(0.0289)	    &	0.6772		(0.0141)	&	0.3978		(0.0137)	\\ 
        GSVM  	&	0.2313	(0.0126)	    &	0.5711		(0.0150)	&	0.2906		(0.0134)	\\ 
        PSVM  	&	0.6332	(0.0141)	    &	0.6537	    (0.0225)	&	0.7349	    (0.0927)	\\ 
        DT  	&	0.2418	(0.0143)	    &	0.4167		(0.0126)	&	0.2915		(0.0112)	\\ 
        RF  	&	0.1808	(0.0120)	    &	0.3022		(0.0120)	&	0.2273		(0.0097)	\\ 
        KNN   	&	0.2241	(0.0131)	    &	0.5271		(0.0148)	&	0.2464		(0.0106)	\\ 
        LR      &	0.6183	(0.0233)	    &	0.6282		(0.0150)	&	0.7449		(0.1371)	\\ \bottomrule 
\end{tabular}
\end{table}

Note that for two reasons we avoided doing KNN classification with different distances like in previous experiments. The main reason is that our aim here is to compare our methods' performance with state-of-the-art papers' results as a baseline.  Furthermore, since the physical characteristic features are clearly important, yet it is not clear how to integrate them with other distances.  The next reason is that most of the preprocessed trajectories are too long -- even though the trajectories are partitioned by time -- and thus the KNN approach can be very inefficient. We remark that the partitioning method in this experiment, as explained above, differs from the method used in binary classification task for Geolife and T-drive trajectory datasets in Sections \ref{sec: experiments} and \ref{sec: other features}.

\paragraph{Comparison with previous studies.}

To the best of authors' knowledge the best reported misclassification rate is $15.2\%$ \cite{DH2018}, which is achieved by intricately designed convolutional neural network. The average misclassification rate of $11.9\%$ we achieved with our vectorized model, using $v_Q$ featurization plus the above 4 physical features, with Random Forest, outperforms this state-of-the-art result. In addition, notice that our method is very efficient and easy to apply as we use simple featurization and simple out-of-the-box machine learning algorithms. Moreover, we only need to tune a single benign hyperparameter in Random Forest for example, namely, the number of estimators, while in CNN one needs to tune many hyperparameters in addition to the design of a useful architecture. 

To compare more, our method outperforms misclassification rates reported in many other studies like \cite{ETNK2016} with $32.1\%$, \cite{WLJL2017} with $25.9\%$ and \cite{ZLCXM2008} with $25.8\%$. A summary of these misclassification rates is given in Table \ref{table: test-error-summary}. Note that the experimental setup of each study is a bit different (we did our best to match the state-of-the-art results~\cite{DH2018}) but all are trying to infer transportation modes.

\begin{table}[h]
\caption{Test error comparison with previous studies}
\label{table: test-error-summary}
\centering
\begin{tabular}{lc}
\toprule
{\bf Study}                                                     &  {\bf Misclassification Rate}   \\ \midrule 
    Using CNN \cite{ETNK2016}                       	        &   $32.1\%$  \\ 
    Using CNN \cite{WLJL2017}                       	        &   $25.9\%$  \\ 
    Inference plus Decision Tree \cite{ZLCXM2008}	            &   $23.8\%$  \\ 
    Using CNN \cite{DCHR2020}                       	        &   $23.2\%$  \\ 
    Our Model with $v_Q$ vectorization                          &   $18.1\%$ \\
    Our Model with $v_Q^{\varsigma +}$ vectorization            &   $16.4\%$ \\
    Our Model with $MD v_Q^+$ vectorization                     &   $15.4\%$ \\
    Using CNN \cite{DH2018}                         	        &   $15.2\%$ \\
    Our Model with $v_Q^+$ vectorization                        &   $11.9\%$ \\ \bottomrule
\end{tabular}
\end{table}

\section{Discussion} \label{Discussion}

This paper establishes the first comprehensive and large-scale empirical comparison of methods to classify trajectories based on their spatial position.  It compares KNN classifiers using standard distance measures, along with new techniques that first create a vectorized feature representation, and can then apply a wide variety of classifiers.  
Data for trajectories in these experiments mostly comes from human movement, potentially via a vehicle.  On these tasks, the vectorized representations typically performed superior to the KNN-based ones across various distance measures.  Moreover, the Random Forest classifier (as is common in other settings~\cite{XGBoost}) was typically the one that achieved the best or nearly best performance.  

Other experiments considered different data sources including characters (written on a tablet), and other features including physical properties such as speed and acceleration.  In these tasks, other methods sometimes showed better performance, specifically a KNN classifier using edit distance with real penalties~\cite{CN04} performed exceedingly well on the character classification tasks, and linear classifiers on the mode-of-transportation tasks.  
We did not specifically design tasks to take advantage of the distances or embeddings that preserved the orientation of direction of the trajectories (as has been done elsewhere~\cite{PP21}, e.g., by modifying the Pigeons dataset~\cite{meade2005three,mann2011nine,mann2014six}), and there was no significant observed advantage of that class of techniques over ones that simply treated trajectories as geometric objects.  Identifying naturally occurring tasks where this is needed, and performing the empirical study is an intriguing future direction.

Beyond the large empirical study, we introduced a new mechanism to select the landmarks for the vectorized representations.  These are dubbed \emph{mistake-driven} and are inspired by the perceptron algorithm and other active learning paradigms.  At each step one of the most misclassified training data trajectories is identified, and a new landmark is selected, with some randomness, near this trajectory with the goal of being informative towards its prediction.  This approach is mostly effective, but can be noisy since (unlike the perceptron algorithm where a data element serves as a support vector) the selected landmark is not guaranteed to boost the accuracy of the identified trajectory.  As a result, we also introduce a simple voting method to aggregate these mistake-driven collections of landmarks.  This approach generates several landmark sets this way, or just chosen from normal distribution at random, and then classifies each trajectory by a majority vote of the classifiers from each landmark set.  The combination of new methods in the mistake-driven landmarks and voting shows to be especially effective.  Note that it is not obvious how to achieve similar voting-based improvements with the KNN-based classifiers for the various distances since the classifiers themselves are deterministic, and would thus all vote the same way.  

We remark that we investigated (but do not present) several other potential data-driven mechanisms to choose landmarks.  These include deriving the gradient for the position of a landmark and performing various gradient-descent-based approaches.  These appeared to consistently get stuck in local minimum, and this line of attack proved unfruitful.

\paragraph{Correlation among classifiers.}
This paper compared 20 different types of classifiers across 6 main experimental tables, each one may be aggregating several related tasks.  Most of the discussion focused on which classifiers performed the best.  However, many of the classifiers are fairly effective at many of the tasks.  One lingering question is how correlated are the performances of similar classifiers -- if they are clustered, in practice it may be useful to check one technique from each grouping.  This we summarize in a correlation matrix.

\begin{figure}[h] 
\centering
\includegraphics[width=0.95 \textwidth]{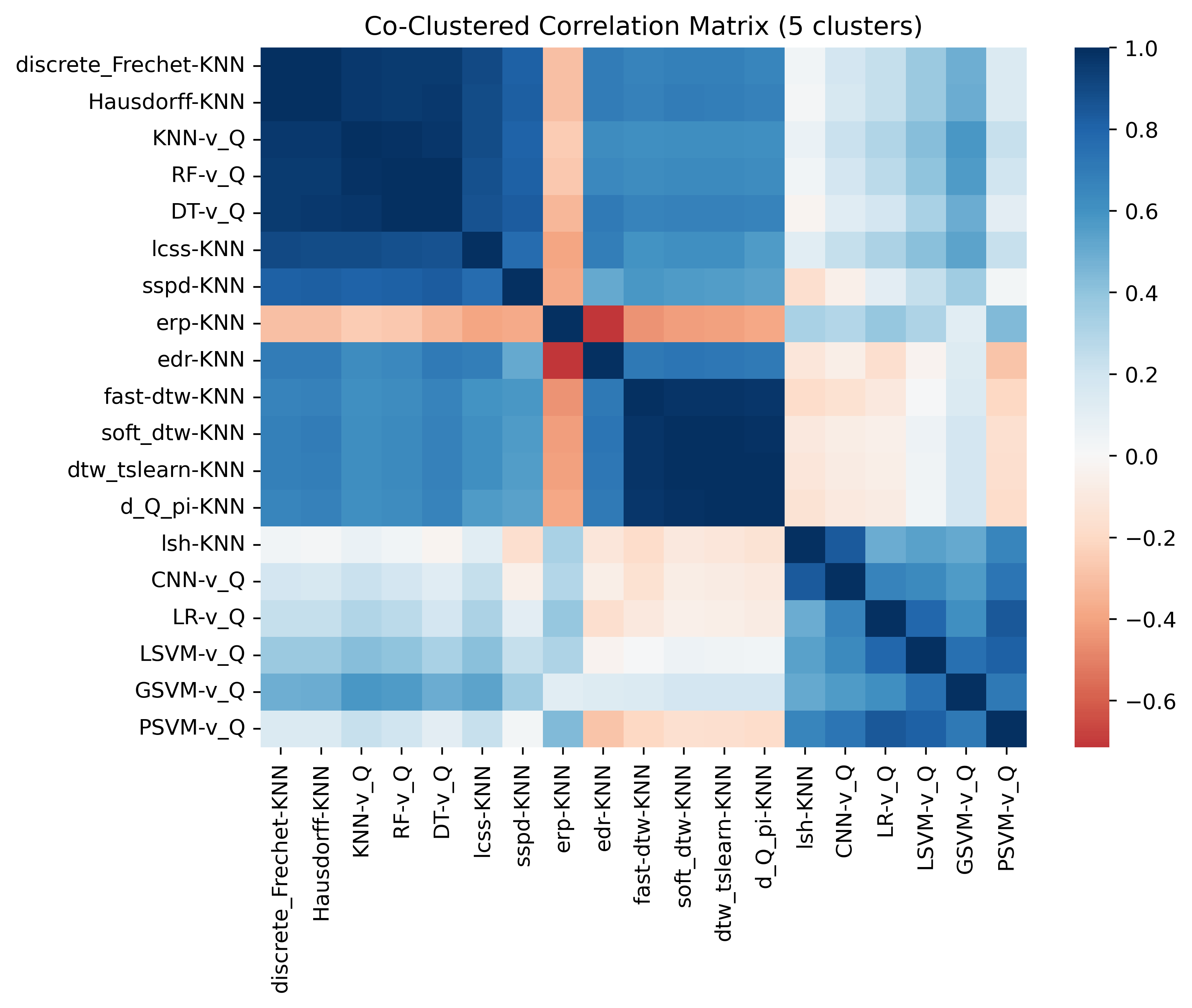}
\caption{Co-clustered correlation matrix for test errors with 19 classifiers.}
\label{fig: correlation} 
\end{figure}

Figure \ref{fig: correlation} shows the co-clustered correlation matrix of misclassification rates calculated for 217 experiments with the 19 classifiers discussed in Section \ref{sec: experiments}.
The correlation is calculated from the 6 main tables, where several experiments are aggregated in the figures.  200 more are included from 100 randomly chosen pairs from Geolife trajectory data, and 100 randomly chosen pairs from T-drive trajectory data.  
The big cluster in the figure indicates that Decision Tree, Random Forest and KNN classifier with $v_Q$-vectorization and KNN with discrete Fr\'echet, Hausdorff, SSPD and LCSS are highly positively correlated. There is another cluster containing all variants of DTW and KNN with $d_Q^{\pi}$. Moreover, all variants of SVM, Logistic Regression and CNN with $v_Q$-vectorization and KNN with LSH have a positive correlation.  The clear outlier is the KNN classifier for edit distance on real penalties (erp-KNN), including negative correlations with all but the vectorized SVM-based methods and CNN.  
The KNN classifier with continuous Fr\'echet  distance is not included in Figure \ref{fig: correlation} as its complexity is high and for most pairs in Geolife and T-drive trajectory data sets it could not be completed within 24 hours (remember that we have chosen more than 200 pairs from these two data sets).

\paragraph{Efficiency.}
In general, the vectorized approaches are significantly more efficient and scalable. 
For instance, training a vectorized classifier on the entire car-bus data set takes between $0.05$ and $0.2$ seconds, or a factor $11$ more (typically under $2$ seconds) with the mistake-driven approach which takes the best of $11$ trials.  In contrast the KNN-based approaches typically took $6$ to $15$ seconds, with only fastdtw (about $1.5$ seconds) and LSH ($0.08$ seconds, using a similar sketch-based approach) nearly as efficient.  
On the Two Persons dataset where the trajectories are larger, the difference is more dramatic.  Verctorized classifiers take $1$ to $2.5$ seconds, with mistake-driven approach again a factor $11$ more.  Where as KNN classifiers typically took $1{,}000$ to $10{,}000$ seconds; with exceptions fastdtw ($32$ seconds) and LSH ($1.5$ seconds).  
And among the KNN classifiers, the most divergent one in terms of performance (erp-KNN) was typically the slowest measured (Fr\'echet was even slower -- more than $80,000$ seconds on Two Persons).

\paragraph{Size of landmarks $\mid \! Q \! \mid$.}

\begin{figure}[h] 
\centering
\includegraphics[width=0.52 \textwidth]{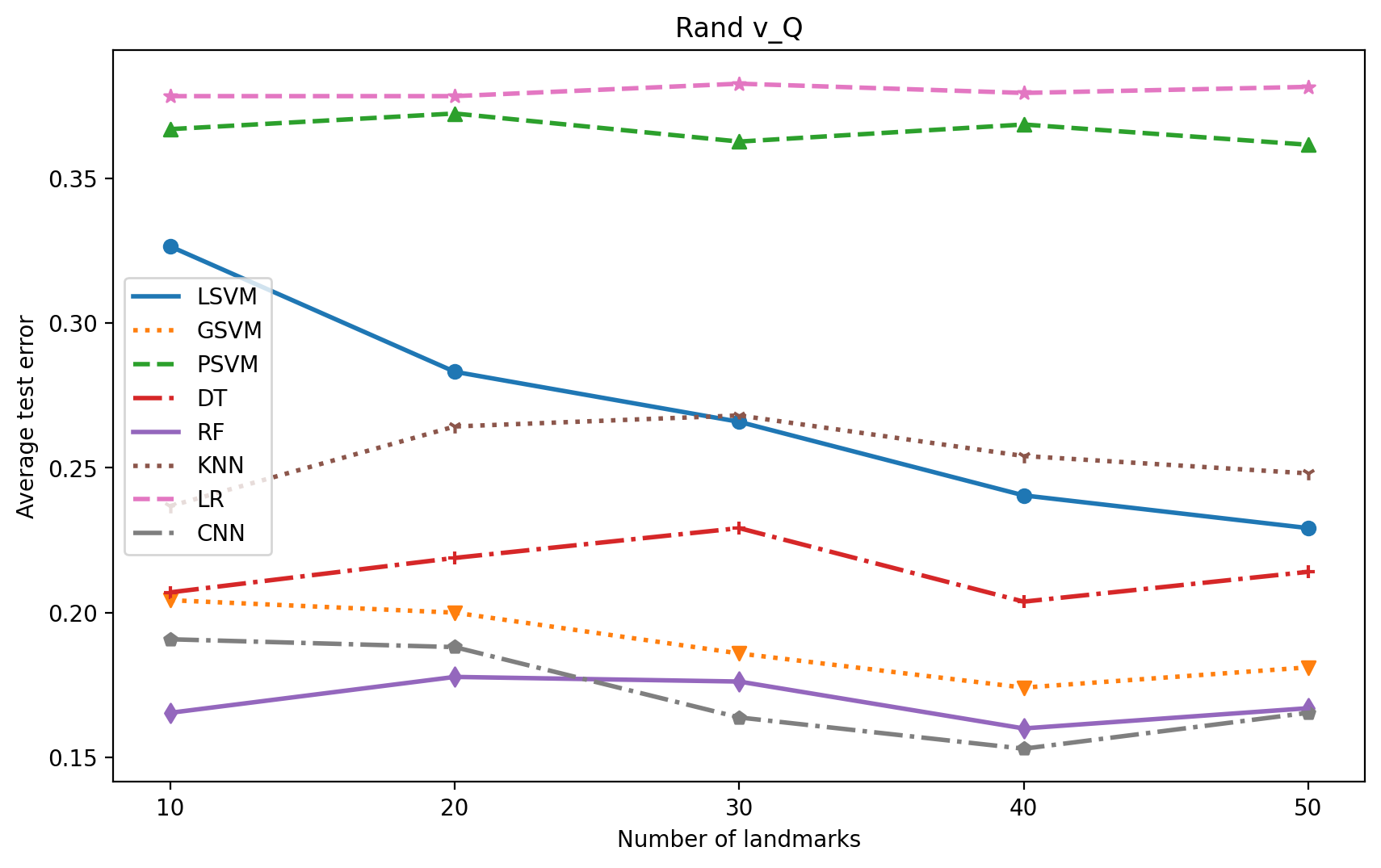}
\caption{Classification test errors of Car-Bus data with 8 classifiers employing $v_Q$-vectorization and different number of random set of landmarks.} 
\label{fig:Rand_v_Q} 
\end{figure}

To choose the right landmark size for experiments, we opted for 20 as we observed 20 landmarks are enough to get a good performance on most datasets. However, we did a simple experiment on the car-bus dataset for different values of $\mid \!\! Q \!\! \mid$ (10, 20, 30, 40 and 50).  As it can be observed, generally speaking, increasing the number of landmarks slightly improves the test errors but kind of flattening at about 20.  As an example, we have given the results with $v_Q$-vectorization in Figure \ref{fig:Rand_v_Q}.

\paragraph{Final Recommendations.}
Ultimately, for the consistent best accuracy, we recommend Random Forest (RF) with Vote(MD $v_Q$) features as a first choice to most likely achieve the best accuracy.  
However, RF with $v_Q$ features offers a very simple, effective, and efficient classifier.  However, one may want to try several, and KNN with ERP is worth also trying.  
Finally, if one has meta data available, or physical characteristics (e.g., length, velocity, acceleration, jerk) make sense for the application, it is recommended to append them to the featurized representation.

\clearpage

\bibliographystyle{plain}
\bibliography{references}

\end{document}